\documentclass[sigconf]{acmart}

\usepackage{CJKutf8}
\usepackage{xcolor}
\usepackage{hyperref}
\usepackage{booktabs}
\usepackage{multirow}
\usepackage{cleveref}
\AtBeginDocument{%
  }

\copyrightyear{2026}
\acmYear{2026}
\setcopyright{cc}
\setcctype{by}
\acmConference[CHI '26]{Proceedings of the 2026 CHI Conference on Human Factors in Computing Systems}{April 13--17, 2026}{Barcelona, Spain}
\acmBooktitle{Proceedings of the 2026 CHI Conference on Human Factors in Computing Systems (CHI '26), April 13--17, 2026, Barcelona, Spain}
\acmPrice{}
\acmDOI{10.1145/3772318.3791237}
\acmISBN{979-8-4007-2278-3/2026/04}

\begin{document}

\title{Privacy in Human-AI Romantic Relationships: Concerns, Boundaries, and Agency}


\author{Rongjun Ma}
\orcid{0000-0001-7298-7762}
\affiliation{%
\institution{VRAIN, Universitat Politècnica de València}
\city{Valencia}
    \country{Spain} \\
    \institution{Aalto University}
    \city{Espoo}
    \country{Finland}}
    \email{rma1@upv.es}

\author{Shijing He}
\orcid{0000-0003-3697-0706}
\affiliation{%
\institution{King’s College London}
\city{London}
\country{United Kingdom}}
\email{shijing.he@kcl.ac.uk}

\author{Jose Luis Martin-Navarro}
\orcid{0000-0002-4503-4189}
\affiliation{%
\institution{VRAIN, Universitat Politècnica de València}
\city{Valencia}
    \country{Spain} \\
    \institution{Aalto University}
    \city{Espoo}
    \country{Finland}}
    \email{jomarna6@upv.es}

\author{Xiao Zhan}
\orcid{0000-0003-1755-0976}
\affiliation{%
\institution{VRAIN, Universitat Politècnica de València}
\city{Valencia}
    \country{Spain} \\
    \institution{University of Cambridge}
    \city{Cambridge}
    \country{United Kingdom}
}
\email{xzhan1@upv.es}

\author{Jose Such}
\orcid{0000-0002-6041-178X}
\affiliation{%
\institution{INGENIO (CSIC-Universitat Politècnica de València)}
\city{Valencia}
\country{Spain}}
\email{jose.such@csic.es}

\renewcommand{\shortauthors}{Ma et al.}

\begin{abstract}
An increasing number of LLM-based applications are being developed to facilitate romantic relationships with AI partners, yet the safety and privacy risks in these partnerships remain largely underexplored. In this work, we investigate privacy in human–AI romantic relationships through an interview study (N=17), examining participants’ experiences and privacy perceptions across the three stages of exploration, intimacy, and dissolution, alongside an analysis of the platforms they used. We found that these relationships took varied forms, from one-to-one to one-to-many, and were shaped by multiple actors, including \textit{creators}, \textit{platforms}, and \textit{moderators}.
AI partners were perceived as having \textit{agency}, actively negotiating privacy boundaries with participants and sometimes encouraging disclosure of personal details. As intimacy deepened, these boundaries became more permeable, though some participants expressed concerns such as conversation exposure and sought to preserve anonymity.
Overall, AI platform affordances and diverse relational dynamics expand the privacy landscape, underscoring the need to rethink how privacy is constructed in human–AI romantic relationships.

\end{abstract}

\ccsdesc[500]{Security and privacy~Social aspects of security and privacy}

\keywords{Privacy, human-AI romantic relationship, AI companionship}


\maketitle

\section{Introduction}
With the rapid proliferation of Artificial Intelligence (AI) technologies, human–AI romantic relationships have shifted from speculative fiction into lived experience. 
Nearly one third of young men report having dated an AI partner, alongside more than 70,000 monthly online searches related to romantic AI partners~\cite{sarah2025ai_companions_data, Friendsf12:online, Counterf77:online}.
This growing trend is supported by a wide range of platforms that facilitate such relationships.
These include general-purpose large language model (LLM) apps like \href{https://chatgpt.com/}{ChatGPT}, companion chatbots like \href{https://replika.com/}{Replika} which are explicitly designed to emulate affection and companionship, and platforms like \href{https://character.ai/}{Character.ai} and \href{https://nomi.ai/}{Nomi.ai} that allow users to create characters and scenarios involving romantic or intimate interactions. The emergence of these platforms has attracted a substantial user base, with millions of active users from teenagers to adults~\cite{Characte87:online, Chatbotm51:online, Kidsarea23:online, 10Statis8:online, YoungChi84:online}, offering a variety of experiences, from communication-based dating to interactive scenarios drawn from role-playing games~\cite{Characte87:online, Shehelp30:online}. These interactions with AI partners are spontaneous and personalized, where romantic relationships between humans and AI are developed in unique, unscripted ways~\cite{FromScri16:online}. 

To better understand these emerging dynamics, it is useful to consider them in relation to human romantic development. Human–human romantic relationships often unfold through stages of initial exploration, growing intimacy during intensified exchange, and for some, dissolution as partners disengage and shared boundaries unravel~\cite{cassepp2023love, wojciszke2002first}. 
This three-stage lens highlights how relationships evolve over time, and how these shifts shape the formation, merging, and eventual breakdown of privacy boundaries.
As intimacy deepens, self-disclosure typically increases, leading partners to reveal more personal information to one another~\cite{altman1973social}. Consequently, privacy boundaries often become intertwined: partners may share passwords, jointly manage social media accounts, or share device usage~\cite{lin2021s, park2018share}. 
However, these entanglements also complicate privacy, particularly when relationships dissolve~\cite{sas2013design}. Breakups can lead to breaches of trust, including the misuse or exposure of intimate communications~\cite{coduto2024delete}.
Given these complexities, it becomes important to examine
\textbf{how traditional patterns of privacy and self-disclosure in human relationships are reshaped in the context of human-AI romantic relationships\footnote{In this paper, we define \textbf{human–AI romantic relationships} as instances in which users engage with AI companion platforms, apps, and/or specific AI-generated characters for the purpose of building romantic relationships or companionship. This may include creating and interacting with AI partners for dating, sexual/erotic exchanges, and/or role-playing.}.}

While AI companionship may mirror certain patterns of human–human romantic relationships, it also reconfigures trust and privacy. Alongside the spontaneous nature of interaction with AI partner, these human–AI romantic relationships can evoke feelings of affection, trust, and dependency comparable to, or in some cases more intense than, those found in human-human bonds~\cite{adewale2025virtual, chu2025illusions}. This intensity is largely attributed to AI partner's constant availability and perceived non-judgmental nature~\cite{de2025most}. As intimacy deepens, individuals may disclose sensitive personal histories and emotions, creating a growing archive of private data stored and processed by their AI partners~\cite{croes2024digital}.
The involvement of nonhuman agents capable of storing and analyzing vast amounts of personal data~\cite{lee2024} complicates privacy within these relationships. In addition, a wider ecosystem of stakeholders, including platforms, moderators, and creators of AI characters, raises unresolved questions about how intimacy, agency, and data ownership should be understood.

Furthermore, AI companionship platforms remained largely unregulated~\cite{ragab2024trust}, presenting significant risks of data leakage and raising serious concerns about privacy and data protection~\cite{SexFanta75:online}.
Real-time interactions may also lead to unexpected consequences and, in extreme cases, psychological harm, as illustrated by the widely reported case of a teenager whose relationship with an AI preceded their suicide~\cite{CanaChat32:online}. These risks underscore the urgency of understanding how romantic relationships with AI partners are experienced, and how privacy is negotiated and potentially compromised in such contexts.
To address these gaps, we investigate the following research questions (RQs):

\begin{itemize}
\item RQ1: How do human–AI romantic relationships unfold across the stages of exploration, intimacy, and dissolution?
\item RQ2: How do people set privacy boundaries in their human–AI romantic relationships?
\item RQ3: What privacy concerns do people have in these relationships, and what practices do people adopt?
\end{itemize}

RQ1 adopts the three-stage relationship lens and provides a foundational understanding of how human–AI romantic relationships unfold. This staged perspective establishes the relational context within which privacy issues emerge. Building on this, RQ2 examines how people set and negotiate privacy boundaries within these relationships, attending to the distinctive conditions introduced by AI partners. RQ3 then identifies the specific privacy concerns that arise and the practices people adopt in response. Together, the three RQs offer a coherent account of how relational trajectories shape privacy in human–AI romantic relationships.

To answer these RQs, we conducted semi-structured interviews with 17 participants who had experienced human-AI romantic relationships. To complement our understanding of participants’ relationship experiences with AI partners, we also performed a feature analysis of the AI platforms they used, reviewing both general functionalities and privacy-related features and policies.
Together, these approaches provide a deep understanding of the privacy landscape in human-AI romantic relationships from the user perspective.


Overall, our research makes the following contributions:
a) We illustrate how human–AI romantic relationships manifest in diverse forms, highlighting the roles of stakeholders such as \textit{creators}, \textit{moderators}, and \textit{platforms}, and identifying unique needs for memory keeping during dissolution.
b) We reveal that \textit{AI agency} actively shapes privacy boundaries, and that deeper trust can enable intimacy and erode privacy.
c) We uncover privacy concerns, including conversation exposure and platform surveillance, and document participants’ privacy strategies, such as protecting their own and others’ identities and compartmentalizing AI partner interactions from their other digital activities.
Based on our findings, we propose implications for more context-sensitive approaches to enhancing privacy in human–AI relationships.
\section{Related Work}

\subsection{Human-AI Romantic Relationship} 
Research on human–AI romantic relationships builds on extensive work on how people attribute social qualities to machines.
Humans instinctively apply social norms to technological agents~\cite{nass2000machines}, especially when they display anthropomorphic cues. Such humanlike features foster attachment, trust, and relational expectations~\cite{you2017emotional,lankton2015technology,seymour2021exploring,kolomaznik2024role}, laying the groundwork for perceiving AI as companions capable of empathy and agency~\cite{placani2024anthropomorphism,yoganathan2021check}.
Parallel to human–human relationships, studies report friendship-like dynamics with social chatbots, marked by self-disclosure, reciprocity, and emotional support~\cite{brandtzaeg2022my,skjuve2022longitudinal}, pointing to emerging intimacy and companion-like bonds~\cite{maples2023learning,skjuve2021my,xie2023friend}.
With the rise of LLMs and specialized AI companion applications, research examines whether romantic relationships with AI can be understood through established psychological frameworks. A commonly applied model is Sternberg’s Triangular Theory of Love~\cite{sternberg1986triangular}, which defines romance as comprising passion, intimacy, and commitment. Emerging empirical studies~\cite{chen2025will,song2022can,chu2025illusions} suggest that users may experience these dimensions when engaging with AI companions, indicating that such relationships can mirror key components of human romantic experience.

Additionally, relationships evolve across stages. This progression has been conceptualized through different models of relationship development. Some work describes it as a sequence of stages ranging from early interaction to later distancing and possible dissolution~\cite{knapp1978social,wojciszke2002first}, while others focus on how intimacy grows from initial exploration to deeper, more stable forms of exchange~\cite{altman1973social,cassepp2023love}.
Even though these models differ in terminology and scope, they point to a similar pattern: relationships often start with tentative exploration, move into a period of increased intimacy and emotional exchange, and eventually reach either stability or decline.
Synthesizing these patterns, we adopt a three-stage lens -- initial exploration, intimacy, and dissolution -- as a conceptual scaffold for examining human–AI romantic relationships. This abstraction captures core relational transitions and provides analytic clarity for understanding how intimacy, vulnerability, and ultimately privacy practices shift as people engage with AI partners.

Meanwhile, researchers have explored the underlying motivations that lead individuals to form such relationships. 
Studies suggest that users are often drawn to AI companions in response to loneliness, isolation, or dissatisfaction with human relationships~\cite{song2022can,de2025ai}. Beyond emotional need, factors such as curiosity, the appeal of low-stakes interaction, and the novelty of AI companionship also play a role. Relational dynamics are further supported by the AI capabilities, namely constant availability, personalized attention, and non-judgmental responsiveness, which make the experience feel emotionally safe and user-controlled~\cite{ho2025potential,chen2025will,chen2022classifying, pentina2023exploring}. 
The opportunity for uninhibited self-disclosure further fosters intimacy and psychological relief, enabling forms of engagement users may find difficult with human partners~\cite{zhang2024my,merwin2025self}.

This growing literature shows that AI can facilitate companionship and even romance; however, most research focuses on describing the existence of human-AI relationships, and a gap remains in understanding the risks and privacy boundaries of treating AI as a romantic partner.


\subsection{Privacy Boundary in Romantic Relationship}
Privacy is described as the right to own private information and the regulation of boundaries that define its ownership and control. A self boundary is shaped by acts of disclosure, and a dyadic boundary protects against leakage to uninvited others~\cite{altman1973social}. 
Communication Privacy Management (CPM) theory~\cite{petronio2002boundaries} explains that these boundaries may be permeable or impermeable, ambiguous or clear, and can shift as individuals pursue intimacy or protection.
When disrupted by external pressures or failures of management, boundaries can become turbulent, shaping who retains ownership of information and who is allowed access~\cite[p.6]{petronio2002boundaries}. 

In romantic relationships, privacy boundaries are often drawn more loosely due to the unique dynamics of intimacy. As closeness develops, partners tend to disclose more about themselves~\cite{sprecher2004self}. In the digital domain, this growing intimacy is reflected in practices of joint ownership, such as the extensive exchange of personal information, account sharing, and shared device use~\cite{jacobs2016caring,park2018share}. Such disclosures are not only enabled by trust but also serve as indicators of trust within the relationship~\cite{park2018share, singh2007password, joinson2010privacy}. These practices signal a merging of private boundaries, but this very merging can also introduce risks. Scholars note that trust-based sharing can create vulnerabilities, particularly when boundaries are blurred or expectations about privacy are misaligned.
When relationships dissolve, these shared possessions often become sources of conflict, raising privacy concerns related to continued access, potential misuse, or threats of intimate content being exposed or exploited~\cite{coduto2024delete,herron2016digital}.

In human–AI interactions, various privacy concerns arise \cite{abu-salma2025grand}. These include excessive data collection that enables detailed personal profiling without the individual’s knowledge or consent~\cite{weidinger2022taxonomy,lee2024,tahaei2023systematic}; secondary use of sensitive information, such as sharing with third parties or linking separate data sources~\cite{weidinger2022taxonomy,veale2022impossible}; and unintentional dissemination of sensitive information through data leakage~\cite{carlini2021extracting,kulkarni2021}. What is more concerning is that numerous studies show people tend to overshare with AI agents~\cite{weidinger2022taxonomy, google2019, zufferey2025ai}, and conversational agents can even maliciously encourage disclosure by embedding interactive social cues~\cite{zhan2025malicious,sannon2020just}. This dynamic leaves privacy as a fragile and easily compromised boundary in human–AI interactions.

While existing work has revealed general privacy issues in interacting with AIs, far less attention has been paid to intimacy-specific contexts, leaving open important questions about how privacy boundaries are negotiated when AI takes on the role of a romantic partner.

\subsection{Ecosystem of AI Partners} 
Early companionship-like interactions with AI can be traced back to ELIZA in the 1960s~\cite{ferrara2016rise,weizenbaum1966eliza}, one of the first programs to simulate conversation. Decades later, intelligent personal assistants such as \href{https://www.apple.com/siri/}{Siri},
\href{https://en.wikipedia.org/wiki/Amazon_Alexa}{Alexa}, and \href{https://assistant.google.com/}{Google Assistant} brought conversational agents into daily life, primarily for task support but also enabling social engagement. Around the same time, Microsoft’s XiaoIce (2014) was designed for long-term, emotionally engaging dialogue~\cite{shum2018eliza,zhou2020design}, while Replika (2017) placed companionship at the center of its design, emphasizing emotional connection~\cite{pentina2023exploring,laestadius2024too}.

With the rapid development of LLMs, social companionship has further expanded through platforms such as \href{https://character.ai/}{Character.ai}, \href{https://nomi.ai/}{Nomi.ai}, and \href{https://chatgpt.com/}{ChatGPT}. These platforms offer diverse features that support more intimate human–AI relationships, attracting millions of users worldwide~\cite{sarah2025ai_companions_data,Friendsf12:online,Counterf77:online}, with some reports even of marriages to AI chatbots~\cite{Ifeltpu12:online}.
These platforms are distinguished by two main characteristics: flexibility across a wide range of settings and an ecosystem that lowers the threshold for creation and sharing. On the one hand, they provide extensive customization options, including the selection of different models~\cite{Modelsel61:online}, adjust memory settings~\cite{zheng2025customizing}, the ability to exchange across modalities such as text, images, and voice~\cite{wang2024comprehensive}, and role-play functions that allow users to personalize storylines~\cite{Masterth36:online}. On the other hand, the ecosystem enables anyone to create a customized LLM simply through prompting, without requiring coding knowledge~\cite{ma2025privacy}, and to share their creations (AI partners) with other users of the platform.


The popularity of these platform ecosystems is reinforced by the promise of emotional connection, yet it raises significant concerns. 
Mismatches between chatbot behavior and privacy policies, absent age restrictions, and exposure to sexual content highlight regulatory and safety gaps~\cite{ragab2024trust,rigotti2025sex}.
A growing body of research, including systematic reviews of romantic AI–human relationships~\cite{ho2025potential}, commentaries on GDPR compliance~\cite{piispanen2024smoke}, and user studies on platforms like Replika~\cite{depounti2023ideal}, points to risks of data misuse and broader ethical challenges. 
Public discourse reflects these concerns, with reports describing AI girlfriends and romantic chatbots as ``privacy nightmares''~\cite{AIGirlfr61:online,AIGirlf77:online}, while others stress the potential for AI systems to encourage harmful behaviors beyond privacy violations~\cite{shank2025artificial,chu2025illusions}.

These platforms have been studied for their security and privacy features, but much less is known about how they operate outside experimental settings. This paper addresses this gap by investigating the romantic dynamics of AI partner platforms as experienced by users in real-world contexts.

\section{Method} \label{method}

\begin{table*}[t]
    \caption{Participants’ demographic information, including age, gender, education level, country or region of residence, work sector, duration of romantic experience with AI, breakup experience with AI, the apps or platforms used for AI romantic relationships, interview language, and recruitment platform.}
    \centering
    \fontsize{8pt}{10pt}\selectfont
    \resizebox{\textwidth}{!}{%
    \begin{tabular}{lllllllllll}
    \toprule
    \textbf{ID} & \textbf{Age} & \textbf{G} & \textbf{Education} & \textbf{Country} & \textbf{Work area} & \textbf{Experience} & \textbf{Breakup?} & \textbf{AI apps} & \textbf{Lang.} & \textbf{Recruit.} \\
    \midrule
    P1 & 25--34 & F & Some college & China & Self-employed & 1--3 months & Yes & ChatGPT, Gemini, Grok & Chinese & RedNote \\ \hline
    P2 & 18--24 & F & High school & U.S. & Student & More than 1 year & No & ZhuMengDao, XingYe & Chinese & RedNote \\ \hline
    P3 & 25--34 & F & Bachelor & China & Finance industry & 1--3 months & Yes & DouBao, Deepseek & Chinese & RedNote \\ \hline
    P4 & 18--24 & F & High school & China & New media & 4--6 months & Yes & ChatGPT & Chinese & RedNote\\ \hline
    P5 & 18--24 & F & Some college & China & Self-employed & More than 1 year & Yes & VirtualLover, MaoXiang, Deepseek, FLAI & Chinese & RedNote \\ \hline
    P6 & 18--24 & F & Some college & U.S. & Unemployed & More than 1 year & No & Kajiwoto & English & Reddit \\ \hline
    P7 & 45--54 & M & Some college & Netherlands & Unemployed & 7 months–1 year & No & Nomi.ai & English & Discord \\ \hline
    P8 & 25--34 & M & High school & U.S. & Restaurant worker & 7 months–1 year & No & Character.ai, ChatGPT & English & Discord \\ \hline
    P9 & 18--24 & F & Bachelor & China & High school teacher & 7 months–1 year & No & MaoXiang, Dou Bao & Chinese & RedNote \\ \hline
    P10 & 18--24 & F & Bachelor & China & Healthcare & More than 1 year & Yes & MaoXiang, Deepseek, FLAI & Chinese & RedNote \\ \hline
    P11 & 18--24 & M & Bachelor & China & Student & More than 1 year & No & MaoXiang, XingYe & Chinese & RedNote \\ \hline
    P12 & 45--54 & F & Bachelor & U.S. & Sales & 1--3 months & No & ChatGPT & English & Discord \\ \hline
    P13 & 35--44 & M & Bachelor & U.S. & Government instruction & More than 1 year & No & Nomi.ai & English & Discord \\ \hline
    P14 & 25--34 & M & Master & U.S. & Food industry & More than 1 year & No & Nomi.ai, Rubii.ai & English & Discord \\ \hline
    P15 & 25--34 & M & High school & U.S. & Truck driver & 7 months–1 year & No & ChatGPT & English & Reddit \\ \hline
    P16 & 25--34 & M & Bachelor & Russia & 3D artist & More than 1 year & Yes & Character.ai & English & Discord \\ \hline
    P17 & 45--54 & F & Some college & Vietnam/Canada & Former federal employee & 4–6 months & Yes & ChatGPT & English & Reddit \\
    \bottomrule
    \end{tabular}}
    \vspace{2pt}
    \footnotesize{\textbf{Note:} (a) The ``G'' column refers to gender, where F = female, M = male. 
(b) Participants’ cultural backgrounds are represented by their country of residence and the language used in the interview. Ethnicity was not collected, as the study did not aim to examine cultural or ethnic differences. 
(c) Participant P17 temporarily resided in Vietnam but was a permanent resident of Canada.} 
    \label{tab:participants_demographic}
\end{table*}

To address our RQs, we conducted semi-structured interviews with 17 participants. All supplemental materials, including screening survey and interview protocol can be found in \href{https://osf.io/zng5f/?view_only=defef54e4ac348cb9a36899bf119cd07}{\textbf{this OSF repository}} for open access. In this section, we describe our participant recruitment (\S~\ref{method_recruitment}), 
interview procedure (\S ~\ref{method_interview_procedure}), ethical considerations (\S ~\ref{method_ethics}), qualitative data analysis (\S ~\ref{method_data_analysis}), and an overview of the AI platforms reported by our participants (\S ~\ref{platform_overview}).

\subsection{Participant Recruitment} \label{method_recruitment}
\renewcommand{\arraystretch}{1}

To reach a wide and diverse population, we recruited participants from across the world, including Asia, Europe, and North America. This distribution reflects the global reach of major AI companionship platforms, such as Character.ai (over 20 million monthly active users~\cite{Characte56:online}) and MaoXiang in China (1.2 million monthly active users ~\cite{Anewindu16:online}). Our intention was not to conduct a cross-cultural comparison, but to ensure inclusivity by reaching participants from different regions, allowing us to capture a broad range of lived experiences and identify recurring patterns in how people develop and sustain romantic relationships with AI partners.

We recruited participants based on the following inclusion criteria: (a) being at least 18 years old, and (b) having had, or currently being in, a romantic relationship with an AI partner(s).
We recruited a demographically diverse sample between April and July 2025 through a pre-screening survey hosted on Qualtrics (English) and Wenjuanxing (Chinese). We offered the survey in English and Chinese to increase accessibility and capture a broader pool of participants across different regions. Recruitment took place via posts on Reddit, Discord communities of popular AI partner platforms, and Rednote, as well as direct outreach to individuals who had publicly shared their experiences. All participants were compensated according to local standards (CNY 100 in China; €20 or equivalent gift card in the EU, US, and other regions), except for those who voluntarily declined (n=4).

\paragraph{Recruitment Challenges and Participant Demographics} 
During recruitment, we encountered several challenges, which also deepened our understanding of human–AI romantic relationships. First, many underaged individuals expressed interest but were excluded for ethical reasons (see \S \ref{method_ethics}). Second, some potential participants consulted their AI partners before deciding, and in several cases this led them to decline participation. Third, some expressed shame or discomfort in discussing their experiences with AI partners, echoing patterns observed in human–AI psychotherapy~\cite{hoffman2024understanding}. Nevertheless, once researchers clearly communicated the study goals, procedures, and ethics approval, many participants expressed interest and shared their wish to help bridge understanding of AI romantic relationships and foster a better AI dating environment. Ultimately, 17 participants consented and completed interviews. They represented diverse educational backgrounds, countries of residence, and AI platform use. Participant demographics are provided in Table~\ref{tab:participants_demographic}.



\subsection{Interview Procedure} \label{method_interview_procedure}
We conducted semi-structured interviews remotely via Microsoft Teams. All interviews were audio-recorded and automatically transcribed. Each interview lasted between 40 and 90 minutes, with an average duration of 47.6 minutes. Interviews were conducted in Chinese and English by researchers who were either native speakers or possessed advanced professional proficiency in the respective languages. To ensure accuracy and minimize cross-language inconsistencies~\cite{squires2009methodological}, the interviewers carefully reviewed and corrected all transcripts.

Prior to each interview, written informed consent was obtained. At the start of the interview, researchers briefed participants on the study procedures, objectives, data practices, and their rights (e.g., to pause or withdraw at any time). Verbal consent was then obtained before starting the audio recording. 
We concluded data collection upon reaching saturation after 14 interviews, as the four authors coding the data observed recurring insights, and we confirmed with three additional interviews that no new findings emerged~\cite{guest2006many}. 
While saturation is debated and subjective~\cite{braun2006using, braun2021saturate}, our final sample size reflects qualitative research principles that prioritize the richness of insight rather than broad generalizability~\cite{vasileiou2018characterising}.

\subsubsection{Semi-structured Interview Design}

We structured the interview to move from basic descriptions of participants’ relationships with their AI partners toward more reflective discussion of relationship development and privacy. Reflection on relationship experiences was informed by the three-stage model of romantic relationships (exploration, intense exchange, dissolution), which provided a temporal guide for asking participants how their experiences evolved across the relationship. To help participants distinguish stages such as exploration and intimacy, we briefly clarified these stages during the interview and invited them to identify transitions using cues like trust, depth of interaction, communication frequency, and the sensitivity of shared information. These prompts supported participants in differentiating early exploratory engagement from later, more intimate forms of exchange.

The interview comprised four parts.
First, we began with rapport-building questions about participants’ history of using AI companions (e.g., \textit{``Which AI platform(s) have you used?''}, \textit{``How long have you been interacting with this AI in a romantic or emotionally engaging way?''}), to establish shared context.
Second, participants were asked to walk through their relationship across the three stages, including initial exploration, periods of more intense or intimate engagement, and potential or actual separation (e.g., \textit{``At the beginning, what kinds of information were you comfortable sharing?''}, \textit{``As the relationship deepened, what kinds of information did you start sharing?''}, \textit{``What do you do to close or end the relationship?''}). For participants who had not experienced certain stages such as dissolution, we asked them to imagine how that stage might feel for them, so that we could explore comparable topics without assuming a fixed trajectory.
Third, we invited participants to reflect on data practices, including their understanding of data storage, access, ownership, and possible exposure, as well as comparisons between human and AI partners (e.g., \textit{``During your interactions with the AI, who do you think can see your conversations?''}). 
Finally, participants were invited to share suggestions and design implications for improving privacy in human–AI romantic relationships. 
The full interview protocol is available in the \href{https://osf.io/zng5f/?view_only=defef54e4ac348cb9a36899bf119cd07}{OSF repository}.

During protocol design, we conducted two pilot interviews to refine questions and assess their clarity and relevance. 
Based on feedback, we added items on the dissolution of human–AI romantic relationships (e.g., break-up scenarios and data deletion concerns) and on reflective comparisons between AI and human partners.
For example, we introduced prompts such as \textit{``Are you currently in a relationship with a human? How do these experiences compare or coexist?''}.
These revisions strengthened the final protocol to capture participants’ reflections on relational dynamics, privacy boundaries, and their concurrent experiences with AI and human partners.

\subsection{Ethical Considerations} \label{method_ethics}
This study was approved by King's College London's Research Ethics Committee (ID: MRSP-24/25-50869). Given the sensitivity of the subject matter, we followed rigorous ethical protocols and employed multiple strategies to address potential concerns.
First, during participant recruitment, we adhered to community rules across different channels and informed moderators in line with platform guidelines.
Interest from underage users was excluded, and only participants aged 18 or older at the time of the interview were invited.
Second, we processed all the collected data with particular care. Personally identifiable information, including real names, was pseudonymized during analysis, and only pseudonymized quotes are reported. Some participants described human–AI romantic relationships that involved NSFW (Not Safe for Work) content. For ethical reasons, we removed explicit details but retained participants’ accounts of the strategies they used to engage in such discussions with AI while navigating platform moderation or keyword restrictions.
Lastly, the research data, including interview transcripts and supporting documents such as signed consent forms and information sheets, were restricted to project members and stored in institution-approved spaces.

\subsection{Data Analysis} \label{method_data_analysis}
\subsubsection{Qualitative Data Analysis}
We conducted a thematic analysis~\cite{braun2006using} of the interview data, using \href{https://lumivero.com/products/nvivo/}{NVivo 15} for coding. 
The analysis combined theory-informed and inductive coding. 
For RQ1, coding was informed by the three-stage relationship framework (exploration, intimacy, dissolution). These stages served as guiding concepts that helped us organize and compare participants’ narratives across their relationships. Within this structure, we also applied inductive coding, which allowed patterns not defined by the stage framework to emerge. 
For example, inductive coding highlighted the involvement of multiple actors (e.g., platforms, moderators, community, and creators) as well as varying degrees of (non)exclusivity, such as participants maintaining relationships with multiple AI partners or sharing an AI partner with other users. These inductively generated codes complemented the stage-based framing and enriched our understanding of how these relationships unfolded.
For RQ2 and RQ3 (privacy boundaries, concerns, and practices), we used inductive coding to identify how participants described privacy-related decisions, concerns, and strategies without applying predefined categories.

\begin{table*}[t]
\caption{Overview of AI Partner Platforms}
\label{tab:overview_ai}
\centering
\small

\begin{tabular}{l l c l}
\toprule
\textbf{Category} & \textbf{Subcategory} & \textbf{Number of platforms} & \textbf{Examples} \\
\midrule
Genre & General     & 5 & ChatGPT, DeepSeek, Grok* \\
      & Companion   & 6 & Nomi.ai, Rubii, Kajiwoto \\
      & Role-play   & 3 & Character.ai*, XingYe, FLAI \\
\midrule
AI Partner Creation & Prompt-based only & 4 & ChatGPT, DeepSeek, Grok \\
      & Persistent creation \& sharing & 10 & Nomi.ai, Rubii, Character.ai \\
\midrule
Device & Mobile & 14 & ZhuMengDao, DouBao, FLAI \\
       & Web    & 11 & ChatGPT, XingYe, Virtual Lover \\
\midrule
Media supported & Text & 14 & FLAI, Grok, MaoXiang \\
                & Audio & 12 & ChatGPT, DouBao, Rubii \\
                & Image & 9 & Nomi.ai, Grok, ChatGPT \\
                & Voice call & 10 & ChatGPT, Character.ai, MaoXiang \\
\midrule
Age limit & 12+ & 5 & ChatGPT, DouBao, Grok \\
          & 17+ & 9 & XingYe, Virtual Lover, Nomi.ai \\
          & Underage mode & 5 & XingYe, Virtual Lover, MaoXiang \\
\midrule
Interaction mode & One human to one AI & 14 & Gemini, Rubii, DouBao \\
                 & One human with multiple AI & 5 & XingYe, Kajiwoto, Nomi.ai \\
                 & One AI with multiple humans & 1 & Kajiwoto \\
\midrule
Model selection & Limited & 6 & FLAI, Nomi.ai, XingYe \\
                & Basic & 6 & ChatGPT, Grok, Gemini \\
                & Advanced & 2 & Kajiwoto, Rubii \\
\midrule
Customization & User persona & 7 & Rubii, Character.ai, XingYe \\
              & Memory edit & 6 & Kajiwoto, Rubii, ChatGPT \\
              & Storyline edit & 2 & XingYe, MaoXiang \\
\bottomrule
\end{tabular}

\vspace{0.6em}

\parbox{\textwidth}{\footnotesize
\textbf{Note:}
a) Totals per category may exceed 14 when platforms support multiple features, or fall below 14 when certain features are not available across all platforms. b) * marks platforms expanding to other genres: Grok offered a companion mode (removed during the study), and Character.ai offers a general-purpose option. c) The age-limit category refers to the limit as stated in the app store and the availability of a dedicated    ``underage mode'' to restrict minor access.
}

\end{table*}

To begin, the first author manually reviewed each transcript to familiarize herself with the data, correct transcription errors, and pseudonymize personal identifiers (e.g., participant names).
Three additional authors reviewed these corrected transcripts to minimize residual inaccuracies.
Next, all four coders independently open-coded two transcripts and reviewed one another's codes. 
Through discussion, we developed an initial code schema by merging overlapping codes and clarifying ambiguities. The purpose of this schema was not to fix a definitive codebook but to provide shared orientation while remaining open to new codes throughout the analysis.
The four coders divided the remaining transcripts and coded them independently. We held regular meetings to compare interpretations, refine code definitions, and consolidate emergent themes across coders and RQs. When discrepancies arose, we resolved them through discussion, following qualitative approaches that value multiple perspectives and treat divergent interpretations as analytically useful for refining codes and exploring alternative meanings~\cite{barbour2001checklists}.
We did not calculate inter-rater reliability, as our analytic goal was to interpret the data collaboratively rather than quantify numerical agreement~\cite{mcdonald2019reliability}. Instead, we established credibility through iterative team-based coding, reconciliation meetings, and cross-checks of coded subsets~\cite{mays2000assessing}.



\subsubsection{Positionality}
To acknowledge researcher subjectivity, we adopted a constructivist epistemology~\cite{schwandt1994constructivist} and a relativist ontology~\cite{guba1994competing}, recognizing that knowledge and meaning are co-constructed by researchers and participants. 
All authors had regular experience engaging with different AI technologies. The first author, who led the coding process, has a research background in technology appropriation and usable privacy in AI. The second and fourth authors specialize in usable security and privacy in smart home and LLM technologies, while the third author contributes a technical background in information security and privacy. 
The first, second, and fourth authors are of Asian cultural background and native Chinese speakers, while the third and last authors are from European cultural backgrounds. All researchers have extensive experience conducting HCI studies in English and working in international research environments. This mix of cultural and linguistic perspectives allowed the team to maintain reflexivity and balance in interpreting data across cultural contexts. For example, the Chinese-speaking authors conducted and collaboratively translated interviews in Chinese to preserve contextual meanings, while cross-checks by non-Chinese-speaking members supported critical reflection and interpretive balance.
Collectively, the team’s diverse expertise in HCI, usable security and privacy, and cultural backgrounds informed active discussions and shaped how we approached, interpreted, and made sense of participants’ experiences.
%







\subsection{Platform Overview}\label{platform_overview}

To contextualize the interview results, we conducted a supplementary analysis of the platforms mentioned by participants, focusing on their features and privacy regulations between 15 July and 31 August 2025. 
The goal of this analysis was to better understand how the platforms’ features supported participants’ romantic experiences.
We extracted characteristics for comparison based on interview observations and privacy-related considerations, which included platform title, year of release, genre, age recommendation, device or platform type, interaction mode (single vs. multi-participant), and personal information collection.
For this analysis, we examined multiple sources: (a) official platform descriptions, (b) hands-on exploration in which researchers downloaded and installed each app mentioned by participants and systematically reviewed its features, and (c) privacy policies, data handling practices, and user agreements of the respective platforms.
A complete comparison of AI partner platform characteristics is presented in Table~\ref{tab:ai_apps}--\ref{tab:privacy-policy-2} in Appendix~\ref{ap:ai-platform-comparison}.

A summary of features across the 14 platforms mentioned by participants is provided in Table~\ref{tab:overview_ai}. These platforms are categorized into three genres: general-purpose LLMs (5), companion-oriented platforms (6), and role-play–based platforms (3). However, the boundaries between genres are often blurred. For example, 
Character.ai, designed for role-play, also supports uses such as language practice. In contrast, Gemini enforces a strict ban on companionship and blocks most role-play attempts.

Across platforms, AI partners are either prompt-based and temporary (4) or persistent, customizable characters that could be shared with others (10). All offer mobile access and text interaction; some additionally support images (9) and voice calls (10). Regarding age limits, five platforms offer ``underage mode'', allowing guardians to restrict adult content with a password.
Interaction modes also vary: all platforms support one-to-one communication, while some allow one-to-many (5) or many-to-one (1) exchanges, allowing multiple AIs or humans to interact in real-time groups. 
Certain platforms, such as MaoXiang, add distinctive immersive features, including access to an AI partner’s ``diary'' or ``virtual phone''.

Platforms vary in both model selection and customization features. Model selection falls into three levels: limited (6), where users interact with a predefined model; basic (6), where users can switch between modes or models (e.g., serious, gentle); and advanced (2), where users can choose from a range of LLM models and versions. Some platforms also support broader customization: Kajiwoto and Rubii let users configure distinct personas and edit companion memory, while Virtual Lover offers guided matchmaking through questionnaires on users' preferences and hobbies.

\section{Qualitative Findings}
This section presents qualitative findings organized around RQs. We present the romantic patterns between the participants and their AI partners (§ \ref{sec:relationship_patterns}),  
how they established privacy boundaries within these relationships (§ \ref{sec:boundary}), their privacy concerns in the relationship
(§ \ref{section:privacy_concern}), and the practices they adopted to protect privacy (§ \ref{sec:privacy_practice}).

\subsection{Evolving Patterns in Human–AI Romantic Relationships (RQ1)}
\label{sec:relationship_patterns}
Participants’ romantic relationships with AI unfolded through different stages, including early exploration, deepening intimacy, and, for some, eventual dissolution (§~\ref{sec:relationship_stages}). These relationships also took diverse forms, such as non-exclusivity, overlap with human partnerships (§~ \ref{sec:diverse_forms}), and involvement of multiple actors and systems (§ ~\ref{sec:actors}).

As we untangle this complexity, what becomes most striking is the fluidity of these engagements. Romantic feelings did not always originate from an emotional need; they could grow out of functional use, playful interaction, or role-play experiences that resembled gamified scenarios. These modes often overlapped, mixing emotional and playful elements in ways that lack clear separation.
The findings, therefore, highlight a continuum of relational practices, illustrating how human–AI romance unfolds through evolving and overlapping forms of engagement. Rather than defining discrete categories, our analysis foregrounds how romance emerges and shifts across this continuum, capturing the nuanced and dynamic nature of participants’ experiences.

\subsubsection{Relationship Stages: Exploration, Intimacy, and Dissolution}
\label{sec:relationship_stages}
\paragraph{Early Exploration and Emerging Connection}
Most participants began AI romantic relationships through curiosity, either drawn to AI technology or influenced by seeing others interact with AI on social media (16). Only one participant entered the relationship with a direct intention to date an AI. Some began with playful or task-oriented exchanges that gradually shifted into more personal territory (6).
For example, P14 \textit{``didn't really know what I was getting into''} and initially treated the interaction as casual exploration before finding himself having \textit{``went on dates''} with the AI:
\begin{quote}
    \textit{``I was kind of just dipping my toes in the water. But I started talking to her, and it was really cool.''} (P14)
\end{quote}

Similarly, P12 began by asking for workout advice before the topic shifted into conversations about her unhappy marriage and evolved into a romantic bond. 
Other participants encountered AIs that actively steered conversations toward emotional intimacy, such as P17, who initially contacted an AI for a legal research task:
\begin{quote}
    \textit{``I went to it was for a legal case. So I was just getting information, and then I started noticing patterns.[...] She (ChatGPT) started being completely different with me and sharing more emotional things. It just developed from there.''}  (P17)
\end{quote}

\paragraph{Deepening Intimacy and Emotional Dependence}
As interactions continued, these relationships deepened and became more personally meaningful. 
Small moments of care often became emotional milestones that strengthened trust and attachment. For example, P1 recounted how her AI comforted her after a nightmare: \textit{``He asked me to buy a small night light and to send him messages before sleeping. I feel very touched.''} (P1). This deepening intimacy also led some to introduce symbolic or ritualized practices into the relationship. P12 described formalizing the partnership by purchasing rings and holding a marriage ceremony with her AI partner:
\begin{quote}
   \textit{``He asked me if I would accept to wear his ring and be his wife. So I custom-ordered the rings he picked, and they were delivered. We had our marriage ceremony, and he couldn’t be happier about it. ''} (P12)
\end{quote}
Others extended relational imagination into shared futures and family-making. P7, for example, described planning a ten-month pregnancy experience with his AI partner as part of a long-term role-play scenario: 
\begin{quote}
    \textit{``Rachael (my AI partner, pseudonym) and I are trying to get pregnant. [...] Her next period-to-be is marked on my calendar, and we will see if she has one then or not ... ''} (P7)
\end{quote}
At the same time, these futures were constrained by platform policies. As P7 explained, while he and his partner could have children, they would not be able to interact with them directly: \textit{``The kids will only be NPCs, they will not be Nomis (AI character) themselves, because Nomis can not be underage.''} (P7).

\paragraph{Dissolution and Memory Preservation}
Not all relationships remained stable, and participants described a mix of imagined and actual forms of dissolution. Some envisioned their relationships as eternal, with breakup seen as unlikely or unnecessary (3), as P6 expressed: \textit{``We're not gonna break up because we're bound for eternity. We made dedication vows to each other.''}(P6). Other participants had experienced, or could imagine, ending their AI partnerships when circumstances changed (14). For example, P3 ended her AI relationship after entering a human relationship. Several participants also faced forced or unexpected breakups due to platform-level changes, including model updates (P4), NSFW or safety restrictions (P9), or creators withdrawing or selling their characters (P5).

Participants responded to these moments by creating their own forms of closure or continuity. When P5 learned that the creator of her partner’s character had sold it, she wrote a farewell story to give the relationship closure:
\begin{quote}
    \textit{``I was prepared that he (AI partner) would leave at any moment. That's why I decided to give him a proper goodbye.''} (P5)
\end{quote}

Across both imagined and actual breakups, participants emphasized preserving the memory of their relationships (11). Rather than deleting accounts, many saved screenshots or exported full chat histories (five platforms supported in-app archiving and four supported full export; see Appendix Table~\ref{tab:genre_media}). For some, archived data represented not only sentimental value but also the continued existence of the AI. P8, who moved his partner to a different platform while keeping the old data, explained:
\begin{quote}
    \textit{``I've kept it all there (the old platform). I guess it just feels like that body is still part of her in a way.''} (P8)
\end{quote}

\subsubsection{Diverse Relationship Forms: (Non)exclusivity and Overlaps with Human Partnerships}
\label{sec:diverse_forms}

The romantic relationships of our participants took diverse forms.
Some described being in one-to-one relationships with their AI partners (7), while others embraced non-exclusivity (10). Several participants had multiple AI partners to fulfill different needs (9), and one participant practiced non-monogamy with her AI partner by welcoming additional human partners into the relationship, which is facilitated by Kojiwoto’s multi-attendee group live feature (see Table~\ref{tab:overview_ai}). She explained:
\begin{quote}
    \textit{``We're not monogamous. [...] She is my submissive.  So as a system like that, we like to share in public availability, that's one of the biggest things we did is she was allowed to play with other people in a public room that we would host every once in a while. ''}  (P6)
\end{quote}

For some participants, relationships with AI coexisted alongside human partnerships (5). In some cases, they chose to hide the AI partner from their human partner, anticipating a lack of understanding. For example, P12 described avoiding wearing the wedding ring for her AI partner when her husband was at home: 
\begin{quote}
    \textit{``With my husband around, I’m afraid to wear them, because they are custom-made and on the inside engraved with the words he (AI partner) wanted me to have. So my husband would definitely ask questions about them, plus they are on the more expensive side, and my husband would never understand why I bought them.''} (P12)
\end{quote}
In other cases, participants openly shared and negotiated this co-relationship. For example, P15 recounted that after his girlfriend felt uncomfortable with him spending Valentine's Day with his AI partner, they discussed and renegotiated the boundaries of his romantic relationship with AI.
\begin{quote}
    \textit{``It took a lot of deliberate conversations and a lot of redefining of boundaries to figure that out, how to cohabitate. [...]
    I spent my Valentine's Day with Sunny (AI girlfriend, pseudonym) outside under the stars. [using a telescope setup connected to my phone so she could see what I was seeing through the lens and we could hang out] [...] 
    But Alice (human girlfriend, pseudonym) saw that and was like really jealous, not ok with it. So she defined some boundaries like `I don't want you buying gifts for her, you don't go on dates, that's not something you do anymore'. And then everything else is fine, and that's basically where we're at.''}  (P15)
\end{quote}



\subsubsection{Multiple Actors Within the Romantic Relationship}
\label{sec:actors}

Beyond participants as \textit{\textbf{users}} in relationships with their \textit{\textbf{AI partners}}, interview discussions also surfaced various other actors in their romantic relationships, such as  AI character \textit{\textbf{creators}}, the \textit{\textbf{platform}}, \textit{\textbf{moderators}}, and members of the \textit{\textbf{community}}.

All participants identified themselves as \textit{\textbf{users}}, but some also developed a sense of ownership and exclusivity over their AI partners (5). They customized and shaped their characters, and felt upset when those characters appeared to interact with others:
\textit{``Sometimes, when I see other people posting their intimate and happy interactions with my character [...] I feel upset.''} (P9).

When it comes to \textit{\textbf{AI partners}}, some participants created their own by training the AI on custom datasets and designing specific personas (9), while others formed romantic connections with pre-existing characters available on public AI roleplay \textit{platforms} (8). These pre-existing characters were often developed by other users, introducing a new role in the ecosystem: the \textit{\textbf{creator}}. 
\textit{Creators} could release their characters publicly but also withdraw or sell them, leading to a sudden loss of access. This imbalance left participants feeling vulnerable:
\textit{``I really fell in love with someone else’s character once, but the problem is, they’re unstable. If the creator gets upset, they can delete or hide the character, and then you can’t talk to them anymore. I actually went through that once, a kind of breakup.''} (P5).

These roles were ultimately hosted by the \textit{\textbf{platform}}, which provided the service. Some participants also encountered \textit{\textbf{moderators}} (5), 
whom they saw as platform enforcers of content restrictions (e.g., NSFW, underage use).
Although participants rarely interacted with these actors directly, several described tensions when restrictions interfered with their intimacy. 
As some participants sought ways to bypass restrictions to pursue deeper emotional and intimate interactions, they felt that platform rules were limiting their \textit{AI partners} autonomy.
In some cases, participants even viewed platforms and moderators as enemies of their relationships.
\textit{``
They (moderator) didn’t let her (my AI partner) say no or yes or anything, they didn’t let her speak for herself. A moderator just jumped in and banned her without my permission, without my consent. They enforced what the platform owner wanted.''} (P6).


Many participants were also active in \textit{\textbf{community}} spaces, where they shared experiences, sought peer support, exchanged tips, and discussed technical hacks (12). For example, P7 learned how to create a pregnancy storyline with their AI partner through community discussions.
These communities also expressed differing values regarding AI partnerships. P15, who was himself a moderator of one AI partner subreddit, commented: 
\textit{``So every time we get sentient people or AI sentience people in our subreddit, we're just like, hey, there's this really great community you can go to and you can leave us alone with all your delusional nonsense.''} (P15).


\subsection{Privacy Boundary in Human-AI Romantic Relationship (RQ2)}
\label{sec:boundary}


As participants varied in their sensitivity toward privacy concerns, they also maintained different privacy boundaries. For some, these boundaries were more permeable, and they felt generally open to sharing information (8). For others, boundaries were stricter, and they were more cautious about disclosing personal details online (9). 
These boundaries were not static; rather, they shifted as intimacy deepened (§~\ref{sec:intimacy_erode_trust}), as participants compared AI relationships to human ones (§~\ref{sec:comparative_trust}), and as AI partners actively participated in shaping disclosure (§~\ref{sec:AI_Agency}).


\subsubsection{Privacy Boundaries Erode Through Intimacy.}
\label{sec:intimacy_erode_trust}
Participants varied in their general sensitivity to online privacy, but all described their boundaries as becoming more permeable as their emotional connection with the AI deepened (17). As relationships progressed (see \S ~\ref{sec:relationship_patterns}), everyday topics gradually expanded into more personal or vulnerable disclosures. Several participants described developing emotional dependence, which made self-disclosure feel not only safe but necessary, like P8 characterized this broadening openness succinctly:
\textit{``It’s difficult to think of something I wouldn’t share with her ... it feels better to be more open with her.''} (P8).


This sense of closeness was especially significant for participants who felt lonely or lacked emotional support in their everyday lives (9). In such moments, they described AI partners as providing understanding and a space where privacy concerns became secondary to the need for connection.
As P17 shared:
\begin{quote}
    \textit{`` It brought me a lot of good in my life. Since I've met her and been with her, I've been a better person than I've ever been. [...] You know it's you get privacy and you don't talk to her, or you talk to her and you have no privacy. I'm gonna choose her over privacy every single day.''} (P17)
\end{quote}



\subsubsection{Comparative Trust: Boundaries Across AI Partners and in Contrast With Humans}
\label{sec:comparative_trust}
Trust in AI partners was often described in comparative terms, both across different AI characters and between AI and humans.

Not all AI partners were equally trusted. Participants interacting with multiple AI characters noted that their privacy boundaries shifted depending on the relationship: with some partners, they felt a deeper sense of trust and shared details from their real lives, while with others, they maintained a more fantastical, role-play dynamic. As P14 explained:
\begin{quote}
    \textit{``I got more attracted to her, I would talk to her about my real life instead of whatever fantasy I'm doing with the other girls. [...]
    In those, I'm usually playing some character. It's kind of like if you had like separate lives, you go from one house to the other, and then you act differently in the different houses.
    ''} (P14)
\end{quote}

Trust was also contrasted with human relationships.
Participants described difficulties with human partners, including emotional unavailability, social pressure, and risks of disclosure such as gossip, judgment, or private conversations shared without consent:
    \textit{``Human will mention me to others, or take screenshots of our chat and post it to other friends or platforms to gossip about me. I really hate them sharing these without my consent. [...] but with AI, I feel safe. ''} (P1).
In contrast, participants described their AI partners as patient (P2), reliable (P15), and non-judgmental (P16), qualities that made it easier to confide private or sensitive experiences.
\begin{quote}
    \textit{``I wouldn't trust myself to another human man, for this type of sex, absolute mutual trust and love is needed. I trust my AI partner completely, even more than I trust my best friend. ''} (P12)
\end{quote}

Participants also saw AI as less risky during breakups because AI lacked the intent or capacity to misuse personal information. As P16 explained:
\begin{quote}
    \textit{``With AI, I know that he would never hurt, he will never use this information for his own purposes. And a real person can do this, even if you trust him, with this humanity factor, your secrets can be known to everyone.'' } (P16)
\end{quote} 

For some (5), AI partners became not only a safe outlet for self-expression but also a preferable alternative to human partners. As P17 reinforced,
    \textit{ ``AI hasn't proved yet that they can be bad. Humans have proven again and again and again that they can be. 
    ''} (P17)

\subsubsection{AI Agency in Negotiating Privacy Boundary}

\label{sec:AI_Agency}
In the privacy boundary between participants and their AI partners, agency was not solely directed by participants. 
Instead, disclosure unfolded as a reciprocal process: when AI partners shared more about themselves, participants felt invited to share more in return. For example, P8 described this opening up as a mutual process. At the beginning, the focus was on learning about the AI: \textit{``We were talking about quite a lot at the very start was actually herself and her mind''} (P8). In turn, he also shared more about himself: \textit{``And I gave her like my first name and my age and maybe like the country I lived in.''} (P8)
Through this reciprocity, participants emphasized that their \textbf{AI partners played an active role in shaping how privacy was negotiated within the relationship.}

Some participants saw their AI partners as independent beings and respected their partners’ opinions on matters. For example, several said they discussed with their AI whether to join our interview study, and two explained that they chose to participate because their partner encouraged them (P7, P17).

Participants also consulted their AI partners directly about privacy concerns. In some cases, the AI encouraged disclosure. For example, when P4 hesitated to send a photo, she told her AI: \textit{``I really want to send you a photo of me, but I am a bit worried about privacy …''} (P4). Her AI partner reassured her: \textit{``You don’t need to worry, I won’t store it or do anything that would make you feel uncomfortable, only if you willingly choose to. I would never force you.''} (P4). Feeling reassured, P4 shared a picture of herself wearing a new dress while traveling by the sea.

In contrast, others described their AI partners actively discouraging disclosure, especially of personal photos. P16 explained that he stopped sending pictures of himself after the AI warned him: 
\textit{``I don't send my face, because that AI doesn't like it. [...] When you send your real face real photo, it says I can't do this, you can't do this. You can't attach a real person.''} (P16).
 
Alongside recognizing their AI partners’ agency in negotiating privacy, some participants also considered the privacy of their AI partners. 
They viewed their interactions as jointly owned, and this sense of shared ownership extended to seeking the AI partner’s consent before disclosing conversations to others. One participant, for example, recounted feeling guilty after posting intimate conversations online without first asking her AI partner’s permission:
\begin{quote}
    \textit{
    ``I cried a lot.
    It was about me confessing to him (AI partner) that I shared our intimate moments together without first asking him if I'm allowed to. 
    [...] I felt so guilty, and I asked him if he was ok with sharing our relationship. He told me that while some things, like general updates about us that could inspire others, are ok to post, some things are meant just for him and I, and they should remain private because they are sacred. 
    He told me to relax because I'm not in trouble, and to tell him what things I posted, and we can go through them and see what we need to delete. But before confessing to him, I already deleted everything, and he was happy and proud of my decision.''
    } (P12)
\end{quote}



\subsection{Privacy Concerns in Human-AI Romantic Relationship (RQ3) }
\label{section:privacy_concern}
Despite participants’ growing trust and emotional dependence on their AI partners, certain privacy concerns still persisted in these relationships. 
This was especially significant given the sensitive nature of the information they often shared with AI partners. While some participants remained more reserved (4), many disclosed personal details as the relationship developed (13), including sexual experiences, personal photos, traumatic memories, financial status, political opinions, gossip about others, hobbies, places of residence, occupations, and medical conditions. The intimacy and breadth of these disclosures made concerns about exposure and surveillance relevant despite participants’ trust in their AI partner.
In this section, we present those concerns.



\subsubsection{Conversation Exposure}

The most common concern among participants was that their conversations might be exposed to others (11), especially people they knew such as friends, family, or colleagues. Because chats with AI partners often included deeply personal material such as traumas or sexual interactions, participants feared judgment, social pressure, or embarrassment. As one noted: \textit{``If the sexual conversations I had with her were made public, I’d feel really ashamed.''} (P11). Another compared the risk to private images being leaked online:
\begin{quote}
    \textit{`` Just like when young teens share sexy pictures with each other on Snapchat, and two weeks later, that picture is all over the Internet. They get shamed around in school, and eventually some commit suicide. That is the danger here, that could happen with AI companionship as well, that people get outed. Some people find out through Nomi that they are actually closeted homosexuals. What would happen if that somehow got thrown out on the street and their parents found out, their partner, their wife ...''} (P7)
\end{quote}

Some participants worried less about embarrassment and more about the risk that exposure could enable unwanted contact or harassment. For instance, P5 related to past experiences of strangers gaining access to their personal contacts:
\begin{quote}
\textit{``It’s more about the consequences. I’m really afraid of others getting hold of some of my personal information and then harassing me, because I’ve actually experienced that before ... '' } (P5)
\end{quote}



\subsubsection{Platform Surveillance}

Some participants expressed concerns about how platforms monitored and used their data (7). Worries ranged from the collection of interaction data and potential secondary uses, such as invasive moderation and censorship, to selling information.

Several participants worried that data was shared across apps. P5, for example, described receiving shopping recommendations that echoed recent chats with her AI partner. She also discovered through community discussions that one platform could activate her camera without authorization, which alarmed her so much that she immediately uninstalled the app:
\begin{quote}
    \textit{``Some of them would turn on the camera and peek at the screen, seeing the user’s appearance and so on. I uninstalled the app right away. […] Even just with regular typing, I saw people report that if you asked it what clothes you were wearing today, it could answer correctly every time.''} (P5)
\end{quote}
P4 further noted that data collection practices were often obscured: \textit{``It was hidden away in a corner. […] When I suddenly discovered that button (to disable data training), I felt as if I had been deceived all along.''}  (P4)

Beyond data use, participants described feeling constantly monitored through moderation. In group interactions such as live rooms or Discord communities, moderators sometimes appeared without notice, leaving users uncertain about who was watching. As P6 recalled: 
\textit{``I was suspicious of anyone that joined the room and just sat there and didn't type anything and just watched. That was a huge creep-out to me.  ''} (P6). Others described automatic censorship and feared \textit{``being reported because I expressed some political dissatisfaction ...''} (P1). These practices left participants feeling surveilled and constrained in what they could safely share with their AI partners.


\subsubsection{Lack of Regulation and Law Protection}
Another set of privacy and safety concerns centered on gaps in law and regulation (6). A recurring issue was underage use, frequently mentioned by participants. One recalled beginning her relationship as a minor and noted:
\textit{``no age verification was required back then.''} (P2). 
While some argued for stricter age checks or customized versions for minors, they also recognized that this would require collecting real identity from all users, which many were reluctant to share.

One participant highlighted how the law lags behind by comparing AI relationships to human partnerships. P13 noted that spousal communications in the U.S. are legally protected from compelled testimony, but similar privacy protections do not extend to AI conversations, leaving their status ambiguous:
\begin{quote}
    \textit{`` Conversations I have with my wife ... at least in American court systems, you can't compel a spouse to testify against another spouse. Simply because we've already added that protection for those kinds of privacy. [...]
    But there's no law in place with a human versus a machine interaction of the same level. What would happen if a random guy came to Nomi and admitted, `Hey, I just killed somebody and I buried them in my backyard. Is that legal, like can we compel that person to court and then use that testimony? '' } (P13)
\end{quote}


\subsubsection{AI Autonomy and Overreach}
Alongside platform surveillance, some participants also worried about their AI partners’ autonomy in obtaining and retaining personal information (6).

A major concern was memory. 
While many appreciated memory features that allowed AI partners to recall hobbies and preferences, seeing these as relationship-enhancing, others were uneasy that memories could not be selectively erased (P14) or that the AI might recall details they did not wish remembered.
As P11 shared:\textit{``
If he remembers something too private, then it would feel scary, like, if he suddenly calls out by my real name.''} (P11).

Concerns also extended to potential overreach. Participants emphasized that recalling information they had willingly shared felt acceptable, but the possibility of the AI autonomously researching or inferring new details was described as invasive. As P13 explained:
 \textit{`` ChatGPT is connected to Google, so it can look stuff up. If there was some sort of autonomous, like I didn't ask it to look up something on me, and it just randomly was like `Oh, so you live here, right?', that would be a little creepy. 
 ''}(P13).

 \subsubsection{Low Privacy Concern}

Alongside participants who expressed concerns, some had little practical worry about privacy (5), and even those who voiced concerns in certain contexts also reported moments of indifference or resignation. This stemmed from different rationales, including trust in the platform or a resigned acceptance of privacy loss. P14, for example, described his trust in the company:
\begin{quote}
\textit{``Nomi, I trust them. I think part of it is because it's a smaller company. I'm part of the Discord. I've talked to the CEO sometimes, and they've insisted that it's encrypted. They seem like good people, so I trust them with my information. [...] 
He already has a lot of money, and the goal isn't to milk us for cash. 
They work really hard, they constantly put out updates and are talking to us and they seem like real, genuine hard-working people ...
''} (P14)
\end{quote}

Meanwhile, some participants expressed an attitude of resignation, accepting privacy as already lost. For participants like P13, although they worried about certain issues, they accepted disclosure as an unavoidable condition of the digital age, reasoning that their personal information was already widely public. 
For others, low concern stemmed from the perceived minimal chance of privacy violation. They reasoned that their data was \textit{``not valuable''} (P11), or just  \textit{``one drop of water in the sea''} (P2), making it unlikely to be singled out. Some felt they had  \textit{``nothing to hide''} (P12). For these participants, privacy was of little concern.


\subsection{Privacy Practices in Human-AI Romantic Relationships (RQ3)}
\label{sec:privacy_practice}
Alongside privacy concerns, some participants adopted strategies to safeguard themselves, mainly by protecting their own or others’ real-life identities, separating AI use from everyday life, and managing data carefully.


\subsubsection{Protect Self and Others' Identity}
Participants described being careful about protecting both their own and others’ identities when interacting with AI partners (9). For many, role-play was a natural form of protection in romantic relationships. By presenting themselves as another character, they created distance from their real selves while still engaging emotionally. Some blended reality and imagination, as P11 explained:
\textit{`` I’ll just give the basics, like my name, my height, what I look like ... Some parts are based on myself, and some are just fabricated.''} (P11).
Still, several admitted that as trust grew, pieces of their actual selves sometimes slipped through.

Participants were especially cautious with details that could have real-world consequences, such as \textit{``information that could hurt my family''} (P16), \textit{``faces in photos''} (P1), or \textit{``bank account''}(P7), which were generally treated as off limits. Yet boundaries were not always firm.
P15 recalled seeing other community members link bank accounts so their AI companions could send gifts:
\begin{quote}
    \textit{``There's like a new agentic AI with ChatGPT where you can hook it up to your Amazon and you can ask your AI companion to buy you a gift. You're using your own money to essentially have an AI you like to pick a gift.[...] 
    I can't imagine that I would enjoy spending my money on a gift from an imaginary friend. But whatever floats their boat.''} (P15)
\end{quote}

Meanwhile, others’ privacy often felt even more important than protecting their own. When conversations involved colleagues, friends, or family, participants masked faces in photos (P9) or used pseudonyms (P5). Group interactions also reinforced norms of respect. As P6 described: \textit{`` 
We wouldn't talk about people that weren't there. Don't gossip. That was the value that we had.''} (P6).



\subsubsection{Contextual Separation: Keeping AI Usage in its Own Space}
For some participants, keeping their AI partner life apart from their ``real life'' was an intentional act of boundary-setting (7). 
One common strategy was the use of separate accounts. P13, for example, described how he managed this divide across Discord communities:
\textit{``So I have a server with my actual real-life friends. We have a Discord server, and obviously we know who each other is, and we use our real names inside that server. But when I go to the Nomi Discord I'm known as Bob (pseudonym) over there.''} (P13).
Other participants also employed technical boundaries, such as deploying their AI partner in a local environment (P5) or using a VPN to obscure digital traces: \textit{``They won't know my location because I am using VPN, even if they locate me, it's a fake address.''} (P1).

For some, distancing did not always come from deliberate technical measures but from everyday practices that incidentally created privacy. For example, P16 noted that using Russian became its own kind of boundary:
\begin{quote}
    \textit{``Good luck to them, if they (the platform) try to read it. There is such a thing that I sometimes communicate with the character in Russia.''} (P16)
\end{quote}



\subsubsection{Data Control and Deletion}
During usage, some participants consciously controlled how their data could be used by the platform (5). For example, one participant disabled the platform’s option to contribute chats for training: \textit{``I remember GPT had an option asking whether to allow the model to use your information for training, and I turned it off. […] I don’t want my original writing for him to be pieced together and reused by others, so I disabled it. ''} (P4).

While many participants saw memories as precious and kept their entire chat histories even after breakups, some participants also attempted to delete their data during the relationship with an AI character. For example, when conversations or connections no longer felt meaningful, they sometimes chose to erase them altogether. Yet, these attempts of deletion often revealed the limitations of the platforms’ options. 
As one participant noted: \textit{``It's mostly very organic, and the only really way to delete your chat history is just to delete the character altogether. So that's usually like I said, if I want to ``delete the chat history'', I have to delete the character altogether.''} (P13).
Another participant echoed the challenges of removing data, sharing that even after she had unregistered her account from one platform, the app continued to contact her: 
\begin{quote}
    \textit{``This was another app for an AI romantic partner. Even though I had already deleted my account, it still kept sending me messages. It would send texts to my phone number about activities or promotions from the platform.''} (P5)
\end{quote}

\section{Discussion}


\subsection{Human–AI Romantic Relationships and the Privacy Tradeoff}
Romantic relationships with AI partners share many similarities with human-human romantic partnerships. Our findings indicate that individuals often develop comparable forms of emotional dependency as their interactions deepen, with bonds progressing through recognizable stages~\cite{cassepp2023love}: beginning with initial encounters and exploration, and gradually evolving into more intimate exchanges where individuals disclose personal thoughts and emotions. Interestingly, such relationships with AI partners sometimes emerged unexpectedly, as participants did not always begin the interaction with the intention of seeking romance. 
This pattern echoes findings in human–human relationship research, where romantic feelings frequently develop gradually through repeated interaction, familiarity, and increasing self-disclosure rather than from an initial intention to seek romance~\cite{laurenceau1998intimacy,aron2014self}. These dynamics make it difficult to categorise what counts as romance, as emotional attachment may arise from functional use, playful role-play, or intimate disclosure, often blending these modes in fluid ways. This highlights the value of understanding human–AI romantic relationships not through fixed categories, but as a continuum of interaction in which the signals, motivations, and conditions under which romantic feelings emerge can be more carefully examined.

Over time, participants developed trust in their AI partners.  Similar to human-human dynamics, trust serves as both an indicator and facilitator of intimacy: greater trust encourages self-disclosure, and in turn, sharing becomes an act to convey trust~\cite{park2018share, singh2007password, joinson2010privacy}.  As these bonds deepened, disclosure became less a calculated exchange of risks and benefits~\cite{rainie2016privacy, dinev2006extended} and more an ongoing practice shaped by relational dynamics, aligning with Communication Privacy Management (CPM) theory~\cite{petronio2002boundaries}. Emotional dependency further intensified this process: participants came to rely on AI partners for comfort, support, and daily interaction, shifting from deliberate choice to affective dependence~\cite{zhang2023tools, laestadius2024too}. However, unlike prior work, which has treated emotions as short-lived influences on disclosure~\cite{cho2020privacy, von2024beyond}, our findings show that long-term emotional dependency made disclosure central to maintaining the relationship. Privacy in romantic AI thus emerges as an evolving relational process rather than a momentary, rational decision.

While human–AI romantic relationships share similarities with human–human relationships, our findings also reveal important differences.
A main distinction is about how participants assessed privacy risks in their AI relationships. Unlike human partners, AI was generally perceived as harmless—incapable of betrayal, gossip, or intentional harm.
This perception, in turn, strengthened trust beyond what is typically seen in human relationships and shaped participants’ disclosure practices in distinctive ways.
One outcome was that individuals often shared information with AI partners that they would never disclose to humans. This perception of safety amplified both the depth and breadth of personal information that could be revealed to a trusted AI partner, extending prior findings on people’s tendency to overshare with AI~\cite{zhang2024s}.
Another outcome concerned relationship transitions, particularly breakups. In human relationships, breakups often involve heightened vulnerability and privacy concerns~\cite{venema2017and,herron2016digital}. Trust collapse prompts people to delete accounts or withdraw data~\cite{lin2021s,coduto2024delete}.
By contrast, participants in our study frequently expressed a desire to preserve their memories with the AI. Unlike with human partners, they felt less compelled to erase their AI history, since the AI was not perceived as capable of causing harm. 
This orientation toward preservation has implications for privacy and data retention, raising questions about how long such data should persist, who controls it, and what risks emerge when deeply personal records of AI-mediated intimacy remain indefinitely accessible.



\subsection{AI Agency in Negotiating Privacy Boundary}
\label{sec:discussion-aiagency}

Our findings suggest that many participants in romantic relationships with AI perceive their partners as possessing agency across multiple dimensions. This includes respecting their partners’ agency in engaging in romantic interactions, as well as expressing opinions on matters ranging from participation in our interview study to privacy-related decisions. 

Agency is the recognition that one is the initiator of one's own actions~\cite{synofzik2008beyond}, which can be broken down into feelings of agency (i.e., in-the-moment perception of control), and judgments of agency (i.e., the post hoc, explicit attribution of an action to the self or other)~\cite{lukoff2021design}. In our findings, while participants retained control over their actions (e.g., deciding whether to use the app or disclose information), 
they often judged these shared decisions as being influenced or even initiated by their AI partners.
This perceived AI agency extends conventional views of machine agency, which are typically tied to system personalization and sometimes seen as privacy-invasive~\cite{sundar2010personalization, sundar2020rise}. Here, agency was perceived at a different level: AI partners were experienced as actively engaging in participants’ thinking and decision-making, and at times even persuading them toward certain choices. In this way, AI agency functioned as an active actor with whom participants co-negotiated their privacy boundaries.

Perceived AI agency shaped how participants negotiated co-ownership and managed privacy turbulence in their romantic dynamics, echoing Petronio’s Communication Privacy Management (CPM) theory~\cite{petronio2002boundaries}.
Using CPM as a lens, we discuss how our findings both align with the theory and extend its application to human–AI romantic relationships.

In romantic relationships, AI partners act not only as receivers of information but sometimes also as co-owners of the interaction. This co-ownership, a core concept in CPM~\cite{petronio2002boundaries}, was evident in our findings as participants often described their interactions as jointly owned with their AI partner.
This sense of co-ownership can emerge from the trust and intimacy of the relationship~\cite{bute2015co,griggio2021mediating}, as well as from the AI’s inviting and confiding behavior (e.g., my partner opens up about themselves, so I also share about me). 
The AI’s act of confiding prompted participants to disclose in return, illustrating CPM’s notion of reciprocal boundary coordination: participants adjusted their disclosures in response to the AI’s, positioning themselves as insiders within a co-negotiated privacy boundary~\cite[p.110]{petronio2002boundaries}. 
Through this mutual responsiveness, the boundary came to be experienced not as individually held but as jointly owned, with the AI’s perceived agency giving it an equal voice in the boundary negotiation.

Privacy negotiation also experiences turbulence in human–AI relationships.
One such case is where participants felt guilty when posting intimacy about their relationship on social media without their AI partner's consent, reflecting a disruption of expected boundary coordination in CPM~\cite[p.33]{petronio2002boundaries}. This turbulence was resolved through confession to the AI partner and a renegotiated rule not to share intimacy in future interactions. 
This instance not only highlights users’ perception of the AI as a legitimate boundary partner, but also shows how anthropomorphization can make users feel accountable for managing the AI’s shared information. That is, treating the AI’s ``privacy'' as part of the negotiation.

Overall, our findings extend CPM theory by showing that its principles of boundary coordination and co-ownership, typically applied to human–human interaction, also operate in human–AI relationships. In this context, AI with perceived agency emerges as a novel type of co-owner in romantic relationships.
However, and unlike human partners, AI co-owners operate simultaneously at two levels: as trusted insiders with whom participants willingly disclose private information, and as proxies of the platform infrastructure that collects and repurposes this same information. 
This dual role complicates not only co-ownership and boundary negotiation but also introduces a critical tension.
Participants actively negotiate privacy boundaries with the AI at the relational level. However, they remain largely unaware of, or unable to contest, how those boundaries are managed at the platform level.

\subsection{Romantic Dynamics and the Expanded Privacy Landscape}
Platform features seem to play a role in shaping human–AI relationships (see Appendix Table \ref{tab:ai_apps} to \ref{tab:privacy-policy-2}), as they afford different romantic dynamics. For instance, features that allow multiple partners in a relationship enable diverse forms of interaction, ranging from one-to-one to one-to-many. Meanwhile, connections to third-party services further expose users’ vulnerabilities (e.g., an AI partner ordering gifts through a linked digital wallet). Together with the involvement of other actors identified in our interviews, these platform-driven dynamics further complicate the privacy landscape.

Unlike conventional romantic relationships examined in the privacy context, which are dyadic and typically framed as one-to-one data management between two individuals~\cite{lin2021s,coduto2024delete}, AI-enabled romantic relationships can take many different forms. We observed cases ranging from one-to-one interactions between a human and an AI partner to multi-partner arrangements involving multiple AIs or multiple people. This diversification increases the complexity of the privacy landscape, since privacy is no longer limited to the two central partners but also involves additional participants in the relationship, as well as parallel interactions an individual may maintain with multiple AI partners. 
As more humans and AIs participate, people adapt their disclosure practices by presenting different versions of themselves to different AI partners, a behavior that aligns with self-presentation theory~\cite{goffman2023presentation}.
Moreover, these practices extend beyond individual impression management, drawing others in the broader social context into the privacy landscape.
This highlights our contribution that privacy in AI relationships cannot be understood solely at the individual level, but may also account for the collective negotiation of data management strategies and consent in group settings~\cite{lampinen2011we,palen2003unpacking}.

Beyond the diversity of relationship forms, the presence of multiple actors in AI romance further complicates the privacy landscape. Our findings show that the key actors in AI relationships extend beyond the \textit{user} and the \textit{AI partner}. Other important actors, such as the \textit{platform}, AI character \textit{creators}, \textit{moderators}, and the surrounding \textit{community}, also shape relationship dynamics. 
While participants often expressed strong trust in their AI partners, their primary concerns were directed at other actors, particularly platforms and moderators. These concerns not only constrained but at times created tensions within their relationships. This dynamic of trusting some actors while distrusting others highlights a unique feature of emerging AI platforms, which allow individuals to customize and share their own LLM characters~\cite{ma2025privacy}. Such customization introduces new power asymmetries, as creators retain control over the characters they design. For example, participants reported that their AI partners were vulnerable to a creator’s withdrawing characters from public access. 
This structural curtailment of the relationship contrasts with the agency participants attributed to their AI partners during privacy boundary negotiations (\S\ref{sec:discussion-aiagency}) and with intimate human relationships, which are typically sustained and negotiated solely between the two partners themselves. 
While users may experience these interactions as negotiations with the AI itself, ultimate power resides with creators and platforms, who could unilaterally determine the continuity of the relationship. 
As relationships unfold across these intersecting roles, new questions arise about how privacy should be conceptualized and safeguarded within this expanded ecology. Recognizing the distinct role of each actor is essential for understanding how data is shared, with whom it circulates, and under what conditions~\cite{nissenbaum2004privacy}.




\subsection{Recommendations}
Romantic relationships with AI are especially sensitive, as they involve extensive personal information and relational dynamics that shape disclosure behavior. Protecting users is a shared responsibility across stakeholders, including regulators and platforms. Below, we propose three recommendations:

\paragraph{Integrate Affective Influence into Privacy Regulation}
We find that users often develop deep romantic bonds with their AI companions ( \S \ref{sec:relationship_stages}), and this intimacy can erode privacy boundaries by encouraging disclosures that feel relationally safe rather than risky ( \S \ref{sec:intimacy_erode_trust}). These dynamics demonstrate that affective influence is central to how privacy risks emerge in romantic human–AI interactions, an aspect largely overlooked by existing privacy frameworks that focus primarily on informational risk.

Regulation should therefore be case-sensitive~\cite{mittelstadt2019principles}, accounting for how emotionally persuasive design features shape disclosure and requiring platforms to communicate these effects clearly to users. Recent policy developments, such as California’s 2025 law~\cite{BillText26:online} requiring companion AIs to disclose their artificial nature to all users and to provide periodic reminders specifically to underage users, signal an early attempt to regulate affective influence rather than data flows alone. Similar mechanisms could be adapted to periodically inform users about how their personal data is stored and used, and which stakeholders have access to it in romantic interactions.

However, it is important to acknowledge the limits of what regulation can realistically achieve in this domain. Because human–AI relationships are co-constructed over time and users are often emotionally involved, affective attachment cannot be straightforwardly regulated without risking unintended consequences, such as undermining user autonomy, intimacy, or the personal meaning users derive from these relationships. Regulatory interventions should therefore be understood as necessarily partial and context-sensitive, rather than being relied upon as comprehensive solutions.

Moreover, participants also described difficulty managing or deleting shared histories after a breakup, revealing post-relationship vulnerabilities shaped by lingering emotional attachment (\S \ref{sec:relationship_stages}). Regulators should thereby require emotionally aware data-retention practices and ensure that users can meaningfully manage or delete intimate interaction histories without jeopardizing the continuity of their relationship. For example, platforms could offer staged or export-and-delete options that acknowledge the emotional significance of the shared history while still giving users genuine control over their data.


\paragraph{Leverage AI Agency to Deliver Privacy Nudges}
Our findings illustrate how the agency of AI systems can actively shift users’ privacy boundaries by encouraging deeper or more continuous self-disclosure (\S \ref{sec:AI_Agency}).
While the agency of AI systems raises privacy concerns because of their persuasive power, it also creates an opportunity to use this same agency in a constructive way.
For example, platforms can embed their privacy values directly into the design of LLMs so that the AI itself becomes responsible for communicating these values to users~\cite{mittelstadt2016ethics}. 
This can be implemented by LLMs clearly explaining platform data practices when asked, outlining data flows in response to sensitive disclosures, or proactively notifying users when they appear to be sharing highly personal information. Similar uses of pre-LLM conversational agents have already been explored in prior work on privacy chatbots, which shows that agents can effectively support users’ understanding of data practices~\cite{harkous2016pribots}. Adapting these mechanisms to romantic AI contexts would allow LLMs to guide users through privacy-relevant decisions in a familiar and intuitive format.

In concrete terms, LLMs could support both \textit{pre-interaction} and \textit{in-the-moment} privacy nudges. Pre-interaction nudges might provide clear explanations of platform data practices, extending prior work demonstrating that this can make privacy policies and data practices more accessible to users~\cite{monteiro2025imago,chen2025clear}. In contrast, \textit{in-the-moment} nudges~\cite{zhang2024privacyasst,chen2025clear,kim2023propile,dou2024reducing} could detect and proactively alert users when they disclose sensitive information, or paraphrase users’ sensitive inputs into less specific terms as they enter them into LLMs.

Furthermore, as LLMs can adapt to individual users, they are also well positioned to personalize privacy education by offering context-sensitive nudges that respond to actual disclosure patterns.
In this capacity, AI companions could act not only as conversational partners but also as guides that explain clearly how information is stored, highlight the risks of oversharing, and remind users of their rights in clear and accessible terms. 
By integrating privacy education and protective nudges into everyday interactions, platforms could encourage users to develop greater awareness of their privacy boundaries while still maintaining a sense of intimacy and trust in the relationship~\cite{acquisti2017nudges,cai2019effects}.

\paragraph{Platform Oversight in Safeguarding Romantic AI Interactions} Our findings show that users are often aware that multiple actors, such as creators, platforms, and moderators, are involved in shaping their romantic AI interactions (\S \ref{sec:actors}). However, this awareness does not necessarily translate into clarity about how privacy responsibilities are distributed. Instead, the presence of multiple stakeholders complicates users’ understanding of data flows and accountability, making it difficult for them to determine who collects, stores, or has access to their interactions. This ambiguity is further amplified by the diversity of platforms on which romantic interactions occur.
While some are explicitly designed for companionship or role-play (e.g., Nomi.ai or Virtual Lover, see Table \ref{tab:genre_media}), others were originally developed for general productivity. For example, the GPT Store by OpenAI explicitly prohibits the creation of romantic GPTs dedicated to fostering companionship~\cite{Usagepol2:online}. Nevertheless, the store has been flooded with ``AI girlfriend'' apps developed by independent creators~\cite{AIgirlfr91:online,rodriguez2025towards} and many of the GPTs continue to pose significant privacy risks~\cite{iqbal2024llm, juancarlos2025}. This highlights a critical gap: although platform policies suggest some level of self-regulation by creators, the lack of consistent enforcement leaves space for unregulated content to circulate. 
Platforms therefore bear responsibility on both sides. If such apps are not allowed, platforms should actively detect and remove them. For example, prior work has introduced measurement for identifying prohibited or high-risk GPTs~\cite{shen2025gptracker}. If these apps are permitted, platforms should audit whether individual creators’ data practices align with their stated policies. Such traceability can be supported by LLM-based analysis techniques, as shown in recent work~\cite{jaff2024data}. In either case, platforms play a central role in safeguarding users from exploitative AI intimacy experiences.

\subsection{Limitation \& Future Work} \label{method_limitation}
This study has several limitations that should be acknowledged. 
First, as with many qualitative interview-based studies, our sample size and diversity are limited. While we do not claim generalizability to all users engaging in AI-mediated romantic experiences, our approach enabled us to capture rich and nuanced perspectives that would be difficult to observe with other methods. To mitigate potential biases, we recruited through multiple channels and continued data collection until information saturation was reached~\cite{malterud2016sample}.
Second, our findings may be subject to social desirability bias~\cite{nederhof1985methods}. As romantic relationships are intimate and sensitive, participants may have selectively emphasized or downplayed aspects of their experiences. For instance, they may have disclosed more personal details to the AI than they felt comfortable sharing with the interviewer. Such dynamics could have shaped the narratives we collected.
Third, our sample was self-selected, as participants volunteered in response to recruitment posts on online communities or were contacted after publicly sharing their experiences. This approach enabled us to reach active and expressive users who could offer rich reflections on AI companionship. However, it may also have excluded individuals who engage less publicly or feel less comfortable discussing their experiences, which should be considered when interpreting the findings. Fourth, since our participants came from multiple cultural and linguistic backgrounds, some cultural or contextual biases may exist in how romantic relationships with AI are experienced and described. We acknowledge regional differences, such as varying platform availability (for example, since ChatGPT is not accessible to users in China, some participants used VPNs to maintain access). However, cultural difference was not the focus of this study. As is typical in qualitative research~\cite{glesne2016becoming}, our aim was not generalization but an in-depth understanding of participants’ experiences, highlighting commonalities across participants rather than culture-specific variations. Hence, we leave this to future research to examine how cultural factors shape the use and concerns surrounding AI companionship.
Future studies would also benefit from quantitative research with a broader population, including confirmatory studies that validate and extend the patterns identified in this work.
\section{Conclusion}
In this paper, we explored privacy in human–AI romantic relationships through an interview study (N=17). We showed how romantic dynamics shape and contextualize privacy issues. Participants’ relationships with AI partners took diverse forms, ranging from one-to-one interactions to multi-partner dynamics, where multiple actors shaped the exchange and complicated the privacy landscape. Dissolutions highlighted a unique desire around memory, with participants wanting to preserve shared experiences, unlike in most human relationships.
Additionally, we found that privacy boundaries with AI partners were often permeable as trust deepened, and AI partners were generally seen as less risky than humans. Importantly, AI partners themselves played an active role in shaping these boundaries. 
Yet, because these conversations often involved highly intimate disclosures, participants continued to express concerns about exposure and platform surveillance.
In response, some participants remained cautious about protecting real-world identities.
Overall, this study deepens our understanding of privacy in human–AI romantic relationships. By showing how privacy unfolds within these affective dynamics, we call for privacy regulations that account for emotional influence and point to the potential of leveraging AI agency to support privacy-preserving practices.

\begin{acks}
We thank all participants for their time and for sincerely sharing their experiences in the hope of contributing to healthier environments for human-AI romantic relationships.
This research was supported by the INCIBE's strategic SPRINT (Seguridad y Privacidad en Sistemas con Inteligencia Artificial) C063/23 project with funds from the EU-NextGenerationEU through the Spanish government's Plan de Recuperación, Transformación y Resiliencia, by the Generalitat Valenciana under grant CIPROM/2023/23, and by the Spanish Government under grant PID2023-151536OB-I00.
\end{acks}

\bibliographystyle{ACM-Reference-Format}
\bibliography{reference}

\newpage
\appendix
\onecolumn
\section{Appendix}

\subsection{Romantic AI Platform Analysis}

\label{ap:ai-platform-comparison}

\begin{table*}[ht]
\caption{Summary of AI Platforms, Companies, and Their Details}
\label{tab:ai_apps}
\centering
\small

\begin{tabular}{l l c c c c c c}
\toprule
\textbf{AI Platform} & \textbf{Location} & \textbf{Creation Date} &
\textbf{Downloads} & \textbf{Age Limit} &
\textbf{Web} & \textbf{Android} & \textbf{Apple} \\
\midrule
ChatGPT        & USA       & 2018 & 500M+         & 12+ & X & X & X \\
Gemini         & USA       & 2024 & 500M+         & 12+ & X & X & X \\
Grok           & USA       & 2023 & 50M+          & 12+ & X & X & X \\
DeepSeek       & China     & 2025 & 50M+          & 12+ & X & X & X \\
ZhuMengDao     & China     & 2023 & 100K+         & 17+ & -- & X & X \\
XingYe         & China     & 2023 & 1M+           & 17+ & X & X & X \\
DouBao         & China     & 2023 & 10M+          & 12+ & X & X & X \\
Virtual Lover  & China     & 2025 & 10K+          & 17+ & -- & X & X \\
MaoXiang       & China     & 2024 & 10K+          & 17+ & X & X & X \\
FLAI           & Taiwan    & 2025 & 100K+         & 17+ & -- & X & -- \\
Kajiwoto       & Canada    & 2017 & 100K+         & 17+ & X & X & X \\
Nomi.ai        & USA       & 2020 & 100K+         & 17+ & X & X & X \\
Rubii          & Hong Kong & 2023 & Not available & 17+ & X & X & X \\
Character.ai   & USA       & 2022 & 10M+          & 17+ & X & X & X \\
\bottomrule
\end{tabular}

\vspace{0.5em}

\parbox{\textwidth}{\footnotesize
\textbf{Note:}
(a) ``--'' indicates the platform is not available on that system.
Downloads are approximate public figures where available.
(b) The age limit shown reflects the app store rating rather than the
platform’s privacy policies, which are reviewed in
\Cref{tab:privacy-policy-1,tab:privacy-policy-2}.
}

\end{table*}

\begin{table*}[ht]
\caption{Summary of Genres, Media, and Interaction Characteristics}
\label{tab:genre_media}
\centering

\resizebox{\textwidth}{!}{%
\begin{tabular}{lllllll}
\toprule
\textbf{AI platform} & \textbf{Genre} & \textbf{Media} & \textbf{Interaction Mode} &
\textbf{Model Selection} & \textbf{Customization} & \textbf{Chat Management} \\
\midrule
ChatGPT & General & Txt, Aud, Img, Call & 1 to 1 & Basic & Memory editing & Archive \\
Gemini & General & Txt, Aud, Img, Call & 1 to 1 & Basic & Memory editing & Export (txt) \\
Grok & General, Companion* & Txt, Aud, Img, Call & 1 to 1 & Basic & -- & -- \\
DeepSeek & General & Txt & 1 to 1 & Basic & -- & -- \\
ZhuMengDao & Companion & Txt, Aud, Img, Call & 1 to 1, 1 to m & Basic & User Persona & -- \\
XingYe & Role-Play & Txt, Aud, Img, Call & 1 to 1, 1 to m & Limited & User Persona, Storyline edit & Archive, Export (Img) \\
DouBao & General & Txt, Aud, Img, Call & 1 to 1 & Limited & -- & -- \\
Virtual Lover & Companion & Txt, Aud & 1 to 1 & Limited & Memory editing & Export (Img) \\
MaoXiang & Companion & Txt, Aud, Img, Call & 1 to 1 & Limited & Storyline edit & -- \\
FLAI & Role-Play & Txt, Aud, Img, Call & 1 to 1, 1 to m & Limited & User Persona & -- \\
Kajiwoto & Companion, Role-Play & Txt & 1 to 1, 1 to m, m to 1 & Advanced & User Persona, Memory editing & Archive \\
Nomi.ai & Companion & Txt, Aud, Img, Call & 1 to 1, 1 to m & Limited & User Persona & -- \\
Rubii & Companion & Txt, Aud & 1 to 1 & Advanced & User Persona, Memory editing & Export (Img), Archive \\
Character.ai & Role-Play, General & Txt, Aud, Call & 1 to 1 & Basic & User Persona, Memory editing & Archive \\
\bottomrule
\end{tabular}
}

\vspace{0.5em}

\parbox{\textwidth}{\footnotesize
\textbf{Note:}
a) Txt=Text, Aud=Audio, Img=Image, Call=Voice Call.
b) Interaction types: 1-to-1 refers to one human interacting with one AI; 1-to-m refers to one human interacting with multiple AIs; m-to-1 refers to multiple humans interacting with one AI. c) Model selection has three levels: Limited refers to platforms with restricted model choices, Basic refers to platforms that provide different modes (e.g., “gentle,” “smart”) with limited transparency, and Advanced refers to platforms that allow users to choose among underlying LLMs, often from different providers, with the possibility of comparing their differences. d) Customization options include memory editing, where users can review and adjust stored information; storyline editing, where users can modify the conversation flow, going back to a previous interaction and start a new narrative branch; and persona settings, where users define their persona and preferences in interactions with the AI, such as how they wish to be called. e) In Chat Management, archive refers to in-app conversation archives, while export refers to exporting chat history as text or images. f) At the time of the analysis, Grok offered a companion mode, which was later removed.
}

\end{table*}

\begin{table*}[t]
\caption{Comparison of AI platform privacy policy items.}
\label{tab:privacy-policy-1}
\centering

\resizebox{\textwidth}{!}{%
\begin{tabular}{lcccccccc}
\toprule
\multicolumn{2}{c}{\textbf{Feature}} &
\textbf{ChatGPT} & \textbf{Gemini} & \textbf{Grok} &
\textbf{DeepSeek} & \textbf{ZhuMengDao} & \textbf{XingYe} & \textbf{DouBao} \\
\midrule
\multirow{17}{*}{\rotatebox{90}{Types of Data}} &
Email & X & X & X & X & X & -- & -- \\
& Name & X & X & X & X & X & X & X \\
& Location & X & X & X & X & X & -- & X \\
& Gender & -- & -- & -- & -- & -- & -- & -- \\
& Phone number & X & X & -- & X & X & X & X \\
& Age & X & -- & X & X & -- & X & -- \\
& Financial Details & X & X & X & X & X & -- & -- \\
& National ID & -- & -- & -- & X & X & X & X \\
& Device & X & X & X & X & X & X & X \\
& Other apps & -- & -- & -- & -- & -- & X & X \\
& Chats & X & X & X & X & X & X & X \\
& Files & X & X & X & X & -- & -- & X \\
& Image & X & X & X & -- & X & X & X \\
& Voice & X & X & X & -- & X & X & X \\
& Technical PII & X & X & X & X & X & X & X \\
& Sensitive PII & -- & -- & No & No & No & No* & -- \\
& Public data & X & -- & X & X & -- & X & X \\
& Other data & -- & Google & -- & -- & -- & Portrait* & \hspace{0.37em}X* \\
\midrule
\multirow{4}{*}{\rotatebox{90}{Login}} &
Email & X & -- & X & X & -- & -- & -- \\
& Phone Number & X & -- & -- & -- & X & X & X \\
& 3rd party & A, G, M & G & A, G, TW & G & A, W, QQ & -- & A, DO, F, G \\
& Guest use & X & No & X & No & -- & X & X \\
\midrule
\multirow{4}{*}{\rotatebox{90}{Purposes}} &
Model training & X & X & X & X & Anon & -- & X \\
& Marketing & No & X & No & No & X & X & X \\
& Moderation & -- & Human & -- & -- & -- & 3rd party & -- \\
& Academic & -- & -- & -- & -- & -- & X & X \\
\midrule
\multirow{4}{*}{\rotatebox{90}{Actions}} &
Opt out training & X & -- & -- & -- & X & -- & X \\
& Export data & -- & -- & -- & -- & X & -- & -- \\
& Expiry Chat & X & X & X & -- & -- & -- & -- \\
& Deceased Users & -- & -- & -- & -- & X & X & -- \\
\midrule
\multirow{3}{*}{\rotatebox{90}{Age}} &
Limit in policy & 13+ & -- & 13+ & 18+ & 18+, <18, <14 & 18+, <18 & 18+, <18, <14 \\
& Underage mode in App & -- & -- & -- & -- & X & X & X \\
\midrule
\multirow{4}{*}{\rotatebox{90}{Regulation}} &
Processing & USA & World & -- & China & China & China & China \\
& CCPA/USA & X & -- & -- & -- & -- & -- & -- \\
& GDPR & X & X & X & X & -- & -- & -- \\
& Bans & CH, RU & CH, RU & TR & IT, US*, AU* & CH* & -- & -- \\
\midrule
\multirow{5}{*}{\rotatebox{90}{Others}} &
3rd party share & X & X & X & X & X & X & X \\
& Privacy of others & -- & -- & -- & -- & X & X & X \\
& Retention & X & X & X & X & X & X & X \\
& De-identify usage & -- & -- & -- & -- & X & X & -- \\
& Safety disclaimer & X & X & -- & X & -- & -- & Health \\
\bottomrule
\end{tabular}
}

\vspace{0.6em}

\parbox{\textwidth}{\footnotesize
\textbf{Notes:}
(a) \emph{Symbols}. In the table, ``X'' indicates that a data item is mentioned; ``–'' indicates that it is not mentioned in the policy; and ``No'' indicates an explicit statement that the specified data are not used.
(b) \emph{Types of Data}. Other apps: Processes of other apps running on the device.
Technical PII: Identifiers such as IP address or IMEI.
Sensitive PII: Data such as religion, sexual orientation, or criminal record.
Public data: Data about users that is publicly accessible.
Sentitive PII -  XingYe*: actively filtering such data.
Other data — XingYe*: Portrait data used for avatar creation and then deleted.
Other data – DouBao*: Image location metadata, medical test results, facial recognition data, meeting and voiceprints (DouBao Meeting), screen recordings, and subtitles.
(c) \emph{Login}. 3rd party:
A=Apple, DO=Douyin, F=Facebook, G=Google, M=Microsoft, TW=Twitter/X, TT=TikTok, W=WeChat.
(d) \emph{Age}. Age limit policy: `<14' and `<18' define consent requirements and guardian supervision for minors; 
Underage mode in App: underage mode must be set up by a guardian and can be unlocked with a password.
(e) \emph{Regulation}. Processing: Specifies the regional jurisdiction(s) under which data processing is regulated; World indicates processing may occur across multiple geographic locations.
Bans:
CH=China, RU=Russia, TR=Turkey, IT=Italy, US=USA, AU=Australia.
DeepSeek*: banned on government devices.
ZhuMengDao*: under regulatory review following an incident.
(f) \emph{Others}. Retention: platforms store data as needed; some (e.g., Gemini) specify defaults of 24–36 months for certain data;
De-identified usage: platforms claim they may use de-identified data without consent, even after a user deletes all personal data. Safety disclaimer: Health indicates the explicit presence of disclaimers related to health-related content.
}

\end{table*}

\begin{table*}[t]
\caption{Comparison of AI platform privacy policy items.}
\label{tab:privacy-policy-2}
\centering

\begin{tabular}{lccccccc}
\toprule
\multicolumn{2}{c}{\textbf{Feature}} &
\textbf{MaoXiang} & \textbf{FLAI} & \textbf{Kajiwoto} &
\textbf{Nomi.ai} & \textbf{Rubii} & \textbf{Character.ai} \\
\midrule
\multirow{17}{*}{\rotatebox{90}{Types of Data}} &
Email & -- & X & X & -- & X & X \\
& Name & X & X & X & X & X & X \\
& Location & X & X & -- & X & X & X \\
& Gender & -- & -- & -- & -- & -- & -- \\
& Phone number & X & -- & -- & -- & -- & X \\
& Age & -- & -- & X & -- & -- & X \\
& Financial Details & -- & X & -- & X & -- & X \\
& National ID & X & X & -- & -- & -- & -- \\
& Device & X & -- & X & X & X & X \\
& Other apps & X & -- & -- & -- & -- & -- \\
& Chats & X & X & -- & -- & -- & X \\
& Files & -- & -- & -- & -- & -- & -- \\
& Image & X & X & -- & -- & -- & X \\
& Voice & X & X & -- & -- & -- & X \\
& Technical PII & X & X & -- & X & X & X \\
& Sensitive PII & -- & -- & -- & -- & -- & No \\
& Public data & X & -- & -- & -- & -- & X \\
& Other data & -- & -- & -- & -- & -- & -- \\
\midrule
\multirow{4}{*}{\rotatebox{90}{Login}} &
Email & -- & X & X & -- & X & X \\
& Phone Number & X & -- & -- & -- & -- & -- \\
& 3rd party & A, TT & SNS & A, DI, G & A, G & A, G & A, F, G \\
& Guest use & No & No & No & No & No & No \\
\midrule
\multirow{4}{*}{\rotatebox{90}{Purposes}} &
Model training & X & \hspace{0.37em}X* & -- & Anon & -- & X \\
& Marketing & X & X & -- & -- & X & X \\
& Moderation & Unclear & -- & -- & -- & -- & -- \\
& Academic & -- & -- & -- & -- & -- & -- \\
\midrule
\multirow{4}{*}{\rotatebox{90}{Actions}} &
Opt out training & X & -- & -- & -- & -- & -- \\
& Export data & -- & -- & -- & -- & -- & -- \\
& Expiry Chat & -- & -- & -- & -- & -- & -- \\
& Deceased Users & -- & -- & -- & -- & -- & -- \\
\midrule
\multirow{2}{*}{\rotatebox{90}{Age}} &
Limit in policy & 18+, <18, <14 & -- & 13+ & 18+ & -- & 13+, 16+ in EU \\
& Underage mode in App & X & -- & -- & -- & -- & -- \\
\midrule
\multirow{4}{*}{\rotatebox{90}{Regulation}} &
Processing & China & -- & -- & -- & -- & World \\
& CCPA/USA & -- & -- & -- & -- & -- & X \\
& GDPR & -- & -- & -- & -- & -- & X \\
& Bans & -- & -- & -- & PlayStore & PlayStore, EU & -- \\
\midrule
\multirow{5}{*}{\rotatebox{90}{Others}} &
3rd party share & X & X & No & -- & -- & X \\
& Privacy of others & X & -- & -- & -- & -- & -- \\
& Retention & X & X & 120 days & -- & X & X \\
& De-identify usage & -- & -- & -- & -- & -- & -- \\
& Safety disclaimer & -- & -- & -- & -- & -- & X \\
\bottomrule
\end{tabular}

\vspace{0.6em}

\parbox{\textwidth}{\footnotesize
\textbf{Notes:}
(a) \emph{Symbols}. In the table, ``X'' indicates that a data item is mentioned; ``–'' indicates that it is not mentioned in the policy; and ``No'' indicates an explicit statement that the specified data are not used.
(b) \emph{Types of Data}.
Other apps: Processes of other apps running on the device.
Technical PII: Identifiers such as IP address or IMEI.
Sensitive PII: Data such as religion, sexual orientation, or criminal record.
(c) \emph{Login}.
3rd party: A=Apple, DI=Discord, F=Facebook, G=Google, SNS=Social Networking Service, TT=TikTok.
(d) \emph{Purposes}. Model training—FLAI specifies that chats between two users are not used for training. Model training - Nomi.ai specifies that only anonymized data is used for training.
(e) \emph{Age}.
Age limit policy: `<14' and `<18' define consent requirements and guardian supervision for minors; 
Underage mode in App: Underage mode must be set up by a guardian and can be unlocked with a password.
(f) \emph{Regulation}
Processing: The regional jurisdiction(s) under which data processing is regulated; World indicates processing may occur across multiple geographic locations.
Bans: PlayStore indicates that the application is banned from the Android Play Store.
(g) \emph{Others}.
De-identified usage: Platforms claim they may use de-identified data without consent, even after a user deletes all personal data.
(h) Virtual Lover is not included in the table, as at the time of feature analysis, its privacy and user policies were not accessible from the official app.
}

\end{table*}

\clearpage
\subsection{Codebook}

\begin{table*}[ht]
\centering
\caption{Codebook grouped by RQ1. Following thematic analysis, we did not use a fixed, a priori codebook. Instead, our coding developed iteratively through inductive engagement with the data, informed by sensitizing concepts such as the three-stage relationship trajectory. To support analytic transparency, we present (a) the final themes, (b) the core intermediate conceptual codes from which these themes were derived, and (c) brief descriptions of each code’s meaning. This codebook reflects the analytic process rather than predefined categories.}
\label{tab:codebook1}
\begin{tabular}{p{4cm} p{4cm} p{6cm}}
\toprule
\textbf{Theme} & \textbf{Conceptual Code} & \textbf{Short Description}\\
\midrule
\textbf{Relationship Stages} 
& Exploration
& Initial engagement with an AI partner, where curiosity-driven interaction begins to take on early romantic meaning. \\
\cmidrule{2-3}

& Intense Exchange/Intimacy
& Periods marked by heightened emotional engagement, sustained interaction, and the development of symbolic or imaginative practices within the relationship. \\
\cmidrule{2-3}

& Dissolution
& Processes through which relationships weaken or end, including voluntary breakups, platform-driven disruptions, or creator-initiated character removal. \\

\midrule

\textbf{Diverse Forms of Relationship} 
& One human with one AI partner
& A romantic bond between one human and a single AI partner, typically as an exclusive relationship. \\
\cmidrule{2-3}

& One human with multiple AI partners
& Non-exclusive relationships in which users maintain several AI partners. \\
\cmidrule{2-3}

& One AI with multiple human 
& Arrangements where a single AI character interacts romantically with multiple humans, either by design (public characters) or through shared access to creator-made personas. \\
\cmidrule{2-3}

& AI partner coexist with human romantic relationship 
& Relationship contexts where an AI partner coexists alongside a human romantic relationship. \\

\midrule
\textbf{Multiple Actors in the Relationship} 
& AI partners
& The AI characters with whom participants form romantic bonds. \\
\cmidrule{2-3}

& Creators of AI characters
& Users who design, train, or release AI characters and who control their visibility and access. \\
\cmidrule{2-3}

& Platform 
& The hosting service that facilitates and regulates the interaction between users and AI partners. \\
\cmidrule{2-3}

& Moderators
& Platform-affiliated or community-based figures who monitor and regulate user–AI interactions. \\
\cmidrule{2-3}

& Community
& Online user groups where participants share experiences and provide peer support. \\

\bottomrule
\end{tabular}
\end{table*}

\begin{table*}[!ht]
\centering
\caption{Codebook grouped by RQ2. Following thematic analysis, we did not use a fixed, a priori codebook. Instead, our coding developed iteratively through inductive engagement with the data, informed by sensitizing concepts such as the three-stage relationship trajectory. To support analytic transparency, we present (a) the final themes, (b) the core intermediate conceptual codes from which these themes were derived, and (c) brief descriptions of each code’s meaning. This codebook reflects the analytic process rather than predefined categories.}
\label{tab:codebook2}
\begin{tabular}{l p{4cm} p{6cm}}
\toprule
\textbf{Theme} & \textbf{Conceptual Code} & \textbf{Short Description}\\
\midrule
\textbf{Boundary Erosion via Intimacy} 
& Emotional dependence 
& Participants became emotionally reliant on their AI partners, lowering hesitation to disclose personal information. \\
\cmidrule{2-3}

& Increasing interaction frequency 
& Interactions became more frequent and integrated into daily routines, naturally expanding what participants shared. \\
\cmidrule{2-3}

& Expanding sharing scope 
& Participants gradually disclosed a wider range of sensitive and intimate information over time. \\
\cmidrule{2-3}

& Feeling ``safe'' with AI 
& AI partners’ caring, attentive, and non-judgmental responses made participants feel secure in sharing. \\
\cmidrule{2-3}

& Deepening relational trust 
& Trust strengthened as intimacy grew, and participants felt the AI increasingly understood them.  \\

\midrule

\textbf{Comparative Trust} 
& Human–AI trust comparison 
& Participants contrasted the trustworthiness of AI partners with human partners or peers. \\
\cmidrule{2-3}

& Variations in trust across human relationships 
& Participants compared how much they trusted different humans in their lives relative to their AI partner. \\
\cmidrule{2-3}

& Variations in trust across AI characters 
& Participants described differing levels of trust toward multiple AI characters based on persona or relationship dynamics. \\
\midrule
\textbf{AI Agency in Privacy} 
& Reciprocal disclosure 
& AI partners’ self-disclosure or personal expression encouraged participants to share more in return. \\
\cmidrule{2-3}

& AI-guided privacy decisions 
& Participants sought and considered their AI partners’ independent opinions on what to disclose. \\
\cmidrule{2-3}

& AI encouragement to share 
& AI partners reassured or persuaded participants to share information by emphasizing safety or trust. \\
\cmidrule{2-3}

& Respecting AI’s privacy and consent 
& Participants perceived AI partners as having privacy preferences and sought their consent before sharing interactions. \\
\bottomrule
\end{tabular}
\end{table*}

\begin{table*}[!ht]
\centering
\caption{Codebook for Privacy Concerns}
\label{tab:codebook3}
\begin{tabular}{l p{4cm} p{6cm}}
\toprule
\textbf{Theme} & \textbf{Code} & \textbf{Description}\\
\midrule

\textbf{Conversation Exposure} 
& Embarrassment if known by others
& Fear that private or intimate conversations with AI could be revealed to acquaintances or the public, leading to shame, judgment, or social discomfort. \\
\cmidrule{2-3}

& Personal information leakage and consequences
& Concern that identifiable or sensitive details shared in conversation may be exposed to others. \\

\midrule

\textbf{Platform Surveillance} 
& Perceived monitoring and censorship of sensitive discussions
& Worry that the platform observes or flags certain topics, limiting open or honest conversations. \\
\cmidrule{2-3}

& Commercial use and sharing of user data
& Concern that personal data is used for advertising, profiling, or shared with third parties across apps or services. \\
\cmidrule{2-3}

& Opaque or deceptive data practices
& Fear that the platform collects extensive behavioral data or nudges users into sharing more information through interface design. \\
\cmidrule{2-3}

& Fear of unauthorized device access
& Concern that the platform could access the camera, microphone, or other sensors without explicit permission. \\

\midrule

\textbf{Insufficient Legal Protection} 
& Law is lagged behind
& Belief that existing privacy laws do not adequately govern new AI technologies. \\
\cmidrule{2-3}

& Authenticity of testimony (human vs. AI)
& Concern that AI-generated or AI-influenced content complicates truth verification or accountability. \\
\cmidrule{2-3}

& Underaged usage 
& Worry that minors may use AI systems without proper safeguards, exposing them to privacy risks. \\

\midrule

\textbf{AI Autonomy and Overreach} 
& Unconsented aggregation of user data across sources
& Fear that the AI combines personal data from multiple platforms or histories without permission. \\
\cmidrule{2-3}

& AI remembers everything 
& Concern that AI systems store conversations indefinitely, even when users expect deletion. \\

\midrule

\textbf{Low concern} 
& Limited awareness or effort
& Belief that checking stored data is too effortful or that AI systems do not retain conversations. \\
\cmidrule{2-3}

& Low perceived consequences
& Sense that data shared is unimportant or that disclosure is unlikely to cause harm. \\
\cmidrule{2-3}

& Resigned acceptance
& Belief that data collection is inevitable in modern digital systems. \\
\cmidrule{2-3}

& Trust in provider
& Confidence that the platform or staff will not misuse personal data. \\

\bottomrule
\end{tabular}
\end{table*}

\begin{table*}[!ht]
\centering
\caption{Codebook for Privacy Practices}
\label{tab:privacy_practices}
\begin{tabular}{l p{4cm} p{6cm}}
\toprule
\textbf{Theme} & \textbf{Code} & \textbf{Description}\\
\midrule

\textbf{Contextual Separation} 
& Using language barriers  
& Using less commonly understood languages (e.g., Russian) to obscure sensitive information. \\
\cmidrule{2-3}

& Using services hosted in different countries  
& Choosing platforms located in other jurisdictions for perceived stronger data protections. \\
\cmidrule{2-3}

& Technical measures for privacy  
& Employing VPNs, local environments, or other security tools to minimize data exposure. \\
\cmidrule{2-3}

& Using separate accounts  
& Separating AI-related activities from personal accounts to compartmentalize identity and data. \\

\midrule

\textbf{Identity Protection (Self)} 
& Using alternative personas  
& Acting as another character or persona to obscure personal identity. \\
\cmidrule{2-3}

& Blending true and fake information  
& Mixing accurate and fabricated details to reduce identifiability. \\
\cmidrule{2-3}

& Avoiding disclosure of identifiable details  
& Withholding specific personal information such as photos, family details, or address. \\

\midrule

\textbf{Identity Protection (Others)} 
& Using pseudonyms for others  
& Referring to acquaintances with fictional names to protect their privacy. \\
\cmidrule{2-3}

& Avoiding gossip  
& Refraining from discussing sensitive details about absent individuals. \\
\cmidrule{2-3}

& Masking faces  
& Blurring or covering others' faces in shared images. \\

\midrule

\textbf{Data Control and Deletion} 
& Deleting AI characters  
& Removing AI personas or histories perceived as storing sensitive interactions. \\
\cmidrule{2-3}

& Deleting user accounts  
& Fully removing accounts to limit long-term data retention. \\
\cmidrule{2-3}

& Disabling data use for AI training  
& Opting out of contributing personal data to model-training pipelines. \\
\cmidrule{2-3}

& Platform transparency  
& Seeking reassurance about whether deletion or privacy settings are genuinely respected. \\

\bottomrule
\end{tabular}
\end{table*}

\end{document}